\documentclass[12pt]{article}

\usepackage{setspace}
\usepackage{amsmath,bm}
\usepackage{amsmath,empheq}
\usepackage{amsfonts}
\usepackage{graphicx}
\usepackage{physics}
\usepackage{cases}
\usepackage[font=small]{caption}
\usepackage{subcaption}
\usepackage{amsmath}
\usepackage{enumerate}
\usepackage[table]{xcolor}
\usepackage{verbatim}
\graphicspath{{Figures/}}
\usepackage[numbers]{natbib}
\usepackage{csquotes}
\usepackage{makecell}
\usepackage{rotating}
\usepackage{multirow}
\usepackage{enumitem}
\usepackage{tabularx, booktabs}
\usepackage{textgreek}
\newcolumntype{Y}{>{\centering\arraybackslash}X}
\usepackage{footnote}
\usepackage[bottom]{footmisc}
\makesavenoteenv{tabular}
\makesavenoteenv{table}
\usepackage[export]{adjustbox}
\usepackage{mathrsfs}
\usepackage{color,soul}
\usepackage{booktabs}
\usepackage{floatrow}
\usepackage{tikz}
\usetikzlibrary{decorations.pathreplacing,decorations.pathmorphing}
\usepackage{setspace}
\newfloatcommand{capbtabbox}{table}[][\FBwidth]
\usepackage[toc,page]{appendix}
\usepackage{tikz}
\DeclareRobustCommand\sampleline[1]{%
  \tikz\draw[#1] (0,0) (0,\the\dimexpr\fontdimen22\textfont2\relax)
  -- (2em,\the\dimexpr\fontdimen22\textfont2\relax);%
}
\definecolor{ao(english)}{rgb}{0.0, 0.5, 0.0}
\definecolor{alizarin}{rgb}{0.82, 0.1, 0.26}
\definecolor{aqua}{rgb}{0.0, 0.6, 1.0}
\definecolor{ao}{rgb}{0.0, 0.0, 1.0}

\newcommand{\bs}[1]{\boldsymbol{#1}}
\usepackage{siunitx}
\usepackage{textcomp}
\usepackage{hyperref}
\usepackage[nameinlink]{cleveref}
\usepackage{soul}

\usepackage[inline]{trackchanges}

\DeclareMathOperator*{\argmin}{arg\,min}

\begin{document}

\title{\Large An Implementation of the Finite Element Method in Hybrid Classical/Quantum Computers}

\author{
\large{ Abhishek Arora$^{\ddagger}$\footnote{These authors contributed equally to this work and are ordered alphabetically.} ,
Benjamin M.~Ward$^{\S}$\footnotemark[1] ,
and Caglar Oskay$^{\ddagger\S}$\footnote{Corresponding author address: VU Station B\#351831, 2301 Vanderbilt Place, Nashville, TN 37235. Email: caglar.oskay@vanderbilt.edu}} \\ 
	\\
	{$^\ddagger$\large Department of Civil and Environmental Engineering} \\ 
	{$^\S$\large Department of Mechanical Engineering} \\ 
	{\large Vanderbilt University} \\
	{\large Nashville, TN 37212} 
}
\date{}
\maketitle
\setstretch{1.3}


\abstract{This manuscript presents the Quantum Finite Element Method (Q-FEM) developed for use in noisy intermediate-scale quantum (NISQ) computers and employs the variational quantum linear solver (VQLS) algorithm. The proposed method leverages the classical FEM procedure to perform the unitary decomposition of the stiffness matrix and employs generator functions to design explicit quantum circuits corresponding to the unitaries. Q-FEM keeps the structure of the finite element discretization intact allowing for the use of variable element lengths and material coefficients in FEM discretization. The proposed method is tested on a steady-state heat equation discretized using linear and quadratic shape functions. Numerical verification studies are performed on the IBM QISKIT simulator and it is demonstrated that Q-FEM is effective in converging to the correct solution for a variety of problems and model discretizations, including with different element lengths, variable coefficients, and different boundary conditions. The formalism developed herein is general and can be extended to problems with higher dimensions. However, numerical examples also demonstrate that the number of parameters for the variational ansatz scale exponentially with the number of qubits, and increases the odds of convergence. Moreover, the deterioration of system conditioning with problem size results in barren plateaus and convergence difficulties.

\vspace{.3cm}
}


\vspace{-0.3cm}
\section{Introduction}
Quantum computing is nearing an inflection point in its evolution, as larger and distributed quantum computing devices are becoming available. These advances are beginning to allow for the development of new methodologies and algorithms for solving computational mechanics problems using quantum computers with the hope of achieving problem scales that are not accessible with classical computing. Despite significant progress in the development of efficient quantum error correction algorithms, the available quantum computers remain noisy and are typically referred to as noisy intermediate-scale quantum, or NISQ computers~\cite{preskill2018quantum}. This manuscript explores a finite element method framework for hybrid-classical NISQ computers using the variational quantum linear solver (VQLS) algorithm. 

Evaluation of linearized systems of equations is a key component in most of the computational methods used for solving linear and nonlinear mechanics problems. Many of the approaches that leverage quantum computing so far consist of constructing a linear system of equations using a classical computer and evaluating this linear system by use of a quantum computer or a simulator. In their seminal work, Harrow, Hassidim, and Lloyd developed a quantum linear solver (HHL) algorithm that scales logarithmically with the linear system size and theoretically provides exponential speedup as compared to classical algorithms for systems of equations with certain properties~\cite{harrow2009quantum}. Several improvements have been made to the original HHL algorithm since its inception (see e.g.,~\cite{ambainis2010variable, chakraborty2018power, subacsi2019quantum, wossnig2018quantum, childs2017quantum})--allowing the solution of relatively large systems (e.g., $2^{17}$ equations evaluated in Ref.~\citep{Perelshtein_2022}). The HHL algorithm is suitable for fault-tolerant systems and requires a level of quantum error correction that is not yet available in current NISQ devices. Quantum algorithms for iterative linear solvers offer a potential pathway to solving systems with high condition numbers~\citep{raisuddin2023a,raisuddin2024q}. 

\citet{bravo2023variational} proposed the Variational Quantum Linear Solver (VQLS) algorithm for solving linear systems in NISQ computers. VQLS is a variational quantum algorithm where the problem is evaluated via classical optimization with the associated cost function (and its derivatives) computed in a quantum computer. In VQLS, the matrix of the linear system is represented as a linear combination of unitary matrices. Quantum circuits are then prepared for each unitary matrix and used in cost function evaluations. A key difficulty in using VQLS is finding an efficient way to decompose the matrix~\cite{aaronson2015read}. A general methodology for symmetric real matrices as a sum of various combinations of Pauli's spin matrices was developed in \cite{pesce2021h2zixy}. However, this approach requires solving for unknown coefficients corresponding to combinations of Pauli's spin matrices, which are typically not readily available for general linear systems. \citet{trahan2023variational} recently employed this approach to solve one-dimensional time-independent Poisson's equation and time-dependent heat and wave equations with finite element method using VQLS. The examples considered in their work consist of homogeneous material domains with constant element lengths in FEM discretization.  It was concluded that the circuit depth and the number of unitaries required to decompose the matrix directly affect the time complexity of the VQLS algorithm, resulting in scalability issues for applications in practical linear systems \cite{trahan2023variational}. Variational algorithms were shown to exhibit optimization convergence difficulties with increasing system size due to barren plateaus and poor representation of the solution space by the variational ansatzes~\cite{PhysRevApplied.20.014054,Cerezo_2021,leone2024practical,zhang2022fundamental}, subsequently, attempts have been made to partially address these difficulties (see e.g., \cite{PhysRevResearch.6.023069,PhysRevLett.126.140502,PhysRevA.104.032401}).

Variational quantum algorithms have recently been deployed to solve solid mechanics and structural mechanics problems. \citet{liu2024quantum} integrated variational quantum eigensolver with the finite element method to compute the fundamental natural frequency of a structure. Lu et al.~\cite{lu2023adhoc} solved the 4th-order PDE Euler-Bernoulli beam equation for general boundary conditions with a variational quantum eigensolver method. The best practices for the variational quantum eigensolver have been reviewed in Ref.~\cite{TILLY20221}. The performance of discrete finite elements with VQLS was recently investigated in the context of solving one-dimensional finite element problems in Trahan et al.~\cite{trahan2023variational}. An alternative type of quantum computer to gate-based systems, a quantum annealing machine, has been used to solve a truss problem discretized with finite elements in a box algorithm~\cite{Srivastava_2019}. Additionally, it has been shown that quantum annealing can achieve high solution accuracy for nonlinear elasto-plastic problems in 1D and 2D \cite{nguyen_quantum_2024}. \citet{raisuddin2022a} developed an iterative quantum annealing algorithm for finite element problems. \mbox{\citet{cao2013quantum}} developed a quantum solver by obtaining a scalable quantum circuit from the structure of the Hamiltonian for Poisson's equation. The number of quantum operations and qubits required for $d$ dimensional Poisson's problem scales linearly in $d$ and polylogarithmic scaling in the inverse of error $(1/\epsilon)$. \mbox{\citet{liu2021variational}} solved Poisson's equation with a finite-difference method discretization by finding the ground state of the Hamiltonian using the variational quantum algorithm. They developed an explicit tensor product decomposition for the matrix into $2 \log_2 N +1 $ terms, where $N \cross N$ is the size of the matrix. However, their approach is applicable only to matrices with a specific structure, which arises from equally spaced grid points and homogeneous diffusivity within the domain.

This manuscript presents a new formulation of the finite element method (Q-FEM) based on the VQLS algorithm for its implementation in NISQ computers. The primary novel contribution of the proposed formulation is in its use of the finite element construction procedure to achieve a unitary representation of the global stiffness matrix. The proposed formulation allows for the automated development of efficient circuits needed for computing the VQLS cost function. The proposed formulation has been implemented in the context of steady-state heat equation for elements with linear and quadratic shape functions, spatially varying properties and lengths, and subjected to natural and essential boundary conditions. The performance of the proposed approach was assessed as a function of problem characteristics (e.g., element type, problem size), as well as the variational ansatz used in problem parameterization. The proposed approach achieves a significant efficiency advantage compared with generic matrix decomposition algorithms \cite{trahan2023variational}.

The remainder of this manuscript is organized as follows: Section~\ref{section:VQLS} provides an overview of the VQLS algorithm, as well as the state preparation method, variational ansatzes, and cost function used in this study. Section~\ref{section:FEM_VQLS} discusses the Q-FEM implementation approach, specifically the treatment of the global stiffness matrix. In Section~\ref{section:Poisson_Eqn}, we discuss unitary and quantum circuit construction in the context of one-dimensional steady-state heat equation discretized using linear and quadratic elements. Section~\ref{sec:numerical_results} presents the numerical verification of Q-FEM relative to ansatz selection, discretization characteristics, and boundary conditions. Section~\ref{section:Conclusion} provides  conclusions and directions for future work.  

\vspace{0.3cm} \noindent {\bf{Notation:}} Arbitrary vectors and matrices are represented by boldface lower and upper-case letters respectively, following the classical computational mechanics literature. The matrix-vector product of matrix $\mathbf{A}$ with a vector $\mathbf{b}$ is represented as $\mathbf{A} \mathbf{b}$. We use upper-case letters without boldface for unitary matrices (that represent circuit operations), for instance, the second-order identity matrix is denoted as $I$. The quantum state vector, inner and outer products are denoted following a `bra-ket' notation as described: a normalized quantum state vector is denoted as $| \cdot \rangle$, an un-normalized quantum state vector is represented with a boldface letter such as $\ket{\mathbf{a}}$, while the inner product between two quantum state vectors $| a \rangle$ and $| b \rangle$ is denoted with $ \langle a | b \rangle $, and the outer product between the same states is written as $|a \rangle \langle b|$. This manuscript uses little-endian notation which means that a qubit state such as $|01\rangle$ refers to $q_0$ qubit in $|1\rangle$ state and $q_1$ qubit in $|0\rangle$ state.

\vspace{-0.3cm}

\section{Variational Quantum Linear Solver} \label{section:VQLS}
Variational Quantum Linear Solver (VQLS) is a quantum-classical hybrid algorithm designed for use in noisy intermediate-scale quantum (NISQ) computers \cite{bravo2023variational}. VQLS solves linear matrix systems of the form $\mathbf{K} \mathbf{u} = \mathbf{f}$ requiring: (1) a process to represent the forcing vector $\mathbf{f}$ as a quantum state, (2) a method for decomposing the matrix $\mathbf{K}$ as a linear combination of $L$ unitary matrices, (3) a representation of the unitary matrices as a series of quantum gates, (4) a variational ansatz that effectively parameterizes the problem space such that a classical optimizer can be employed to find the solution, and (5) a cost function, the minimum of which corresponds to the solution.

The VQLS formulation allows for complex-valued linear systems, but we restrict our attention to real-valued systems without losing generality. Let $\mathbf{K}\in \mathbb{R}^{N\times N}$ be a full-rank, real-valued, square, consistent matrix; $\mathbf{u} \in \mathbb{R}^{N}$ and $\mathbf{f} \in \mathbb{R}^{N}$ are respectively the solution and force vectors. Consider the following quantum representations of the solution and the force vectors:
\begin{eqnarray}
    \mathbf{u} &\rightarrow& \| \mathbf{u} \| |u\rangle = \| \mathbf{u} \| \sum_{i=1}^{N} \alpha_{i} |i\rangle, \\
    \mathbf{f} &\rightarrow& \| \mathbf{f} \| |f\rangle = \| \mathbf{f} \| \sum_{i=1}^{N} \beta_{i} |i\rangle,
\end{eqnarray}
%
%
where $|u\rangle$ and $|f\rangle$ denote the corresponding qubit state vectors, $|i\rangle$ denotes the $i^{\mathrm{th}}$ computational basis state, $\alpha_{i}$ and $\beta_{i}$ are amplitudes of the $i^{\mathrm{th}}$ computational basis state, which correspond to the $i^{\mathrm{th}}$ component of the normalized solution and force vectors. For simplicity, system size, $N$, is chosen such that $N = 2^n$, where $n$ is the number of qubits (excluding ancilla qubits used for algorithmic purposes). The problem that the VQLS algorithm solves is as follows:
\begin{equation} \label{VQLS_Problem}
    \mathrm{Find} \, \ket{u} \, \textrm{such that} \, \mathbf{K} \ket{u} \, \textrm{and} \, \ket{f} \hspace{0.1cm} \textrm{are collinear.}
\end{equation}
The magnitude of the solution vector, $||\mathbf{u}||$, is obtained in the post-processing step after the VQLS algorithm produces a converged solution for the qubit state vector, $\ket{u}$. The magnitude can be obtained using the following expression:
\begin{equation} \label{solutionMagnitude}
    ||\mathbf{u}|| = \frac{|| \mathbf{f}||}{ \langle f | \mathbf{K} | u \rangle }, 
\end{equation}
where $||\mathbf{f}||$ is known, and $\langle f | \mathbf{K} | u \rangle$ is stored during the iterations in the VQLS algorithm, as discussed in Section \ref{sec:costfunction}.

\begin{figure}[htb!]
    \includegraphics[width=16 cm]
    {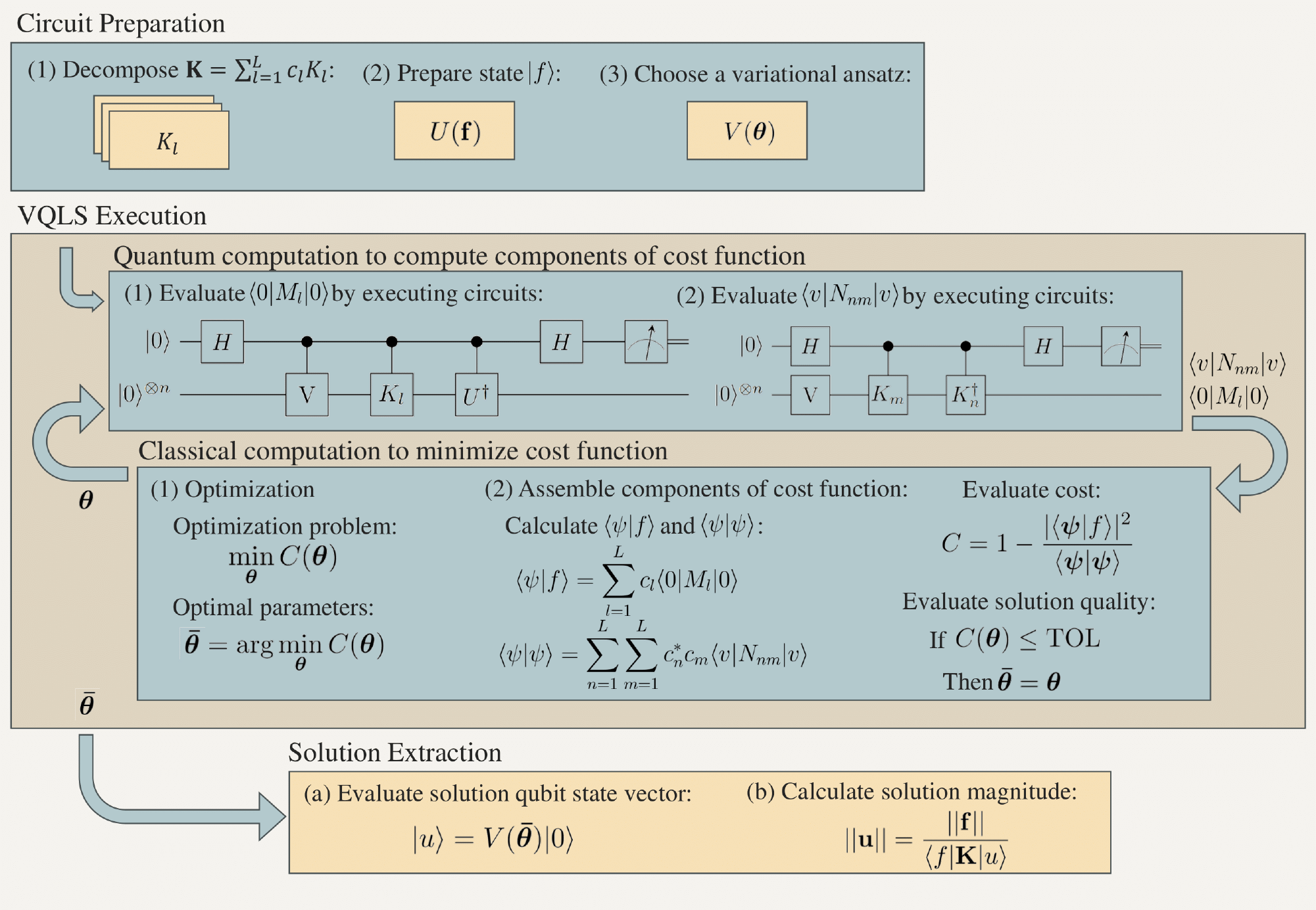}
    \caption{A flow chart for implementation of VQLS.}
    \label{fig:VQLS_Flow}
\end{figure}

The overall strategy of implementing the VQLS algorithm is shown in Fig. \ref{fig:VQLS_Flow} and consists of three steps: preparation of quantum circuits, execution of the VQLS optimization in a hybrid computer, and solution extraction. In the circuit preparation step, $\textbf{K}$ is decomposed into a linear combination of unitary matrices, and a quantum circuit is prepared to represent each unitary matrix. The state vector $|f\rangle$ is prepared using a state preparation algorithm $U(\textbf{f})$. Finally, the circuit representing the chosen variational ansatz is prepared. The variational ansatz defines the parametric space within which the solution of the VQLS system (i.e., $\ket{u}$) is to be searched. In the second step, a classical optimization is performed using a hybrid (quantum-classical) computer to minimize the cost function. The cost function is parameterized by the ansatz parameters, $\bm{\theta}.$ The search space in the variational ansatz is restricted to the quantum states with real probability amplitudes only because our linear system of equations is real-valued. Thus, function evaluations needed to compute the cost as well as its gradients are performed in the quantum computer with the Hadamard test measuring only the real component of the statevector. The cost function is assembled in the classical computer and a search direction is determined by the classical optimization scheme. Optimization iterations continue until a specified tolerance is met. In the third step, the solution parameters $\bar{\bm{\theta}}$ are used to extract the solution unit vector from measurements of the output of the variational ansatz $|u\rangle = V(\bar{\bm{\theta}})$, as well as the solution magnitude using Eq.~\eqref{solutionMagnitude}.

\vspace{-0.3cm} \subsection{State preparation and variational ansatz} \label{sec:state_prep_var_ansatz}

To define a cost function the minimum of which corresponds to the solution state, $\ket{u}$, two operators are defined: 
\begin{subequations} \label{statePreparation}
\begin{eqnarray}
    |v\rangle &=& V(\bs{\theta})|0\rangle, \\
    |f\rangle &=& U(\mathbf{f}) |0\rangle,
\end{eqnarray}
\end{subequations}
in which $V(\bs{\theta})$ is the variational ansatz, and $U$ is the operator that prepares $\ket{f}$. $V$ and $U$ are unitary operators that act on a known reference state, typically taken to be a computational basis state. $\ket{0}$ is chosen as the reference state in Eqs.~\eqref{statePreparation}.

\citet{mottonen2004transformation} proposed a quantum state preparation method using a circuit (i.e., $U$) with a depth of $\mathcal{O}(N)$ and made of controlled rotation gates. Closed-form analytical expressions were employed to calculate the rotation angles. This method was subsequently adapted to incorporate qubit connectivity constraints in real devices~\cite{Shende2006} and to improve efficiency~\cite{Iten_2016, Plesch_2011}. Further advances have been made for special cases, such as symmetry-preserving~\cite{gard_2020} or sparse states~\cite{Feniou2024}. \citet{araujo_2021} and~\citet{Bausch_2022} employed an additional ancilla qubit registry to load information, which traded circuit depth for width. Optimization-based methods using direct machine pulses~\cite{meitei_2021}, genetic algorithms~\cite{creevey_gasp_2023}, or by use of an ansatz~\cite{cerezo_cost_2021} have also been proposed. In this work, we adopt the method by \citet{mottonen2004transformation} because of its general applicability.

The variational ansatz provides the parameterization of the solution space through a sequence of operations: one-qubit rotation gates applied on each qubit modify their state vector representation on the Bloch sphere by rotating around an axis in the real or imaginary plane, and two-qubit entangling gates create superposition between individual qubits and ultimately an entangled all-qubit state. These rotation and entanglement gates are applied in alternating layers. The elements of the parameter vector, $\bs{\theta}$, are the rotation angles for each rotation gate. The variational ansatz is chosen based on the problem characteristics as well as the quantum hardware employed. An ansatz that uses gates native to a particular quantum hardware is termed a hardware-efficient ansatz~\citep{bravo2023variational}, which reduces the gate overhead and avoids additional compiling by hardware controls. The structure of the ansatz itself could be parameterized as explored in~\citep{larose_variational_2019}, whereas a ``fixed-structure ansatz'' employs a constant gate structure--only varying rotation parameters $\bs{\theta}$~\cite{bravo2023variational}. In this work, we tested several fixed-structure anstazes. 

\vspace{-0.3cm} 
\subsection{Cost function} \label{sec:costfunction}

Consider the vector $\ket{\bs{\psi}}$ such that $\ket{\bs{\psi}} (\bs{\theta}) := \mathbf{K} \ket{v}$. A straightforward and intuitive cost function for solving the VQLS problem in Eq.~\eqref{VQLS_Problem} is~\citep{bravo2023variational}:
\begin{equation}
    C_{p} = \langle\bs{\psi}| \mathbf{P} |\bs{\psi}\rangle,
\end{equation}
in which, $\mathbf{P}$ is a projection operator defined as:
\begin{equation}
    \mathbf{P} = I - |f\rangle \langle f|,
\end{equation}
where $I$ is the identity matrix. By this definition of the cost function, it is easy to see that when $\ket{\bs{\psi}}$ is normal to $\ket{f}$, the value of the cost function is $\braket{\bs{\psi}}$, whereas the cost function vanishes when the two vectors are collinear. Normalizing the cost function yields:
\begin{equation}
     \hat{C}_{p} = 1 - \frac{| \langle \bs{\psi}|f\rangle |^2}{\langle\bs{\psi}|\bs{\psi}\rangle}.
\end{equation}
$\hat{C}_{p}$ is defined in terms of the global qubit state. Barren plateaus are observed for such global cost functions in Refs.~\citep{Cerezo_2021,khatri2019quantum}, which motivated the proposal of local cost functions where states or operators are compared with respect to each individual qubit, rather than a global state. Based on the analytical bounds on the variance of the local cost function gradients, it was deduced in Ref.~\mbox{\cite{Cerezo_2021}} that local cost functions exhibit polynomially vanishing gradients, given the ansatz layer depth scales $\mathcal{O}(\log(n))$, compared to exponentially vanishing gradients for global cost functions, which may yield better convergence, especially for large problem sizes. In this work, we primarily use the global cost function, but, we compare the performance of the local cost function with respect to the global cost function for one of our examples in Section \mbox{\ref{section:local_cost_function}}. The local cost function \mbox{\cite{bravo2023variational}} used in this work is shown below:
\begin{eqnarray}\label{eq:local_cost_func}
    \hat{C}_L = 1 - \frac{\langle\psi|U (\frac{1}{n}\sum_{j=1}^{n} |0_j\rangle\langle0_j|\otimes I_{\bar{j}} ) U^\dagger|\psi\rangle}{\langle\psi|\psi\rangle},
\end{eqnarray}
where $|0_j\rangle$ is the zero state on the $j^{\mathrm{th}}$ qubit and $I_{\bar{j}}$ is identity on all qubits except the $j^{\mathrm{th}}$ qubit. This local cost function requires $n$ times as many circuits to be run for the numerator computation, which are notably deeper and hence computationally more expensive than those required by the global cost function. For more details, we refer the interested reader to Ref.~\mbox{\cite{bravo2023variational}}.

The optimization problem to identify the solution to Eq.~\eqref{VQLS_Problem} is then stated as:
\begin{equation}
   \bar{\bs{\theta}} =  \argmin_{\bs{\theta} \in \bs{\Theta} \subset \mathbb{R}^P} \hat{C}_p \left( \bs{\theta} \right),
\end{equation}
such that the best approximation to $\ket{u}$ is obtained as $V(\bar{\bs{\theta}})\ket{0}$. $\ket{u}$ is exact if $\hat{C}_p \left( \bar{\bs{\theta}} \right) = 0$. Note, $\bs{\Theta}$ is the subspace of parameters consisting of $P$ real-valued numbers.

The cost function is primarily computed in a quantum computer. In order to facilitate this, the matrix, $\mathbf{K}$, is decomposed into $L$ unitary matrices:
 \begin{equation} \label{MatrixDecomposition}
      \mathbf{K} = \sum_{l=1}^{L} c_{l} K_{l}.
 \end{equation}
Using Eqs.~\eqref{statePreparation} and~\eqref{MatrixDecomposition}, the components of the cost function are expressed as:
\begin{subequations} \label{costFunctionComponents}
\begin{eqnarray}
    |\langle \bs{\psi}|f\rangle |^2 &=& \sum_{n=1}^{L} \sum_{m=1}^{L} c_{n}^* c_m \langle 0| M_m |0\rangle  \langle 0| M^{\dagger}_n |0\rangle,  \label{numerator} \\ 
    \langle\bs{\psi}|\bs{\psi}\rangle &=& \sum_{n=1}^{L} \sum_{m=1}^{L} c_{n}^*c_{m}\langle v| N_{nm}| v \rangle, \label{denominator}
\end{eqnarray}    
\end{subequations}
where the unitary operators are defined as $M_l=U^{\dagger} K_l V$ and $N_{nm}=K^{\dagger}_n K_m$ and are real valued in our implementation. Consequently, $\langle 0| M_l |0\rangle$ is real, and $\langle 0| M_l |0\rangle =  (\langle 0| M_l |0\rangle)^* = \langle 0| M^{\dagger}_l |0\rangle$. This symmetry is leveraged in the computation of Eq.~\eqref{numerator}. The cost function computation then consists of (1)~preparation of $L(L-1)$ (in case of symmetry, $L(L-1)/2$) circuits associated with $N_{nm}$ and $L$ circuits associated with $M_l$, (2)~employing the Hadamard test to compute $\braket{0}{M_l|0}$ and $\langle v| N_{nm}|v\rangle$ in the quantum computer, and (3)~evaluation of the multiplication and sum operations in Eqs.~\eqref{costFunctionComponents} using a classical computer. The variational ansatz is integrated with the unitaries obtained from the matrix decomposition in Eq.~\mbox{\eqref{MatrixDecomposition}} in the quantum circuit to compute $\braket{0}{M_l|0}$ and $\langle v| N_{nm}|v\rangle$.

\begin{figure}[hbt!]
    \centering
    \includegraphics[width=12cm]{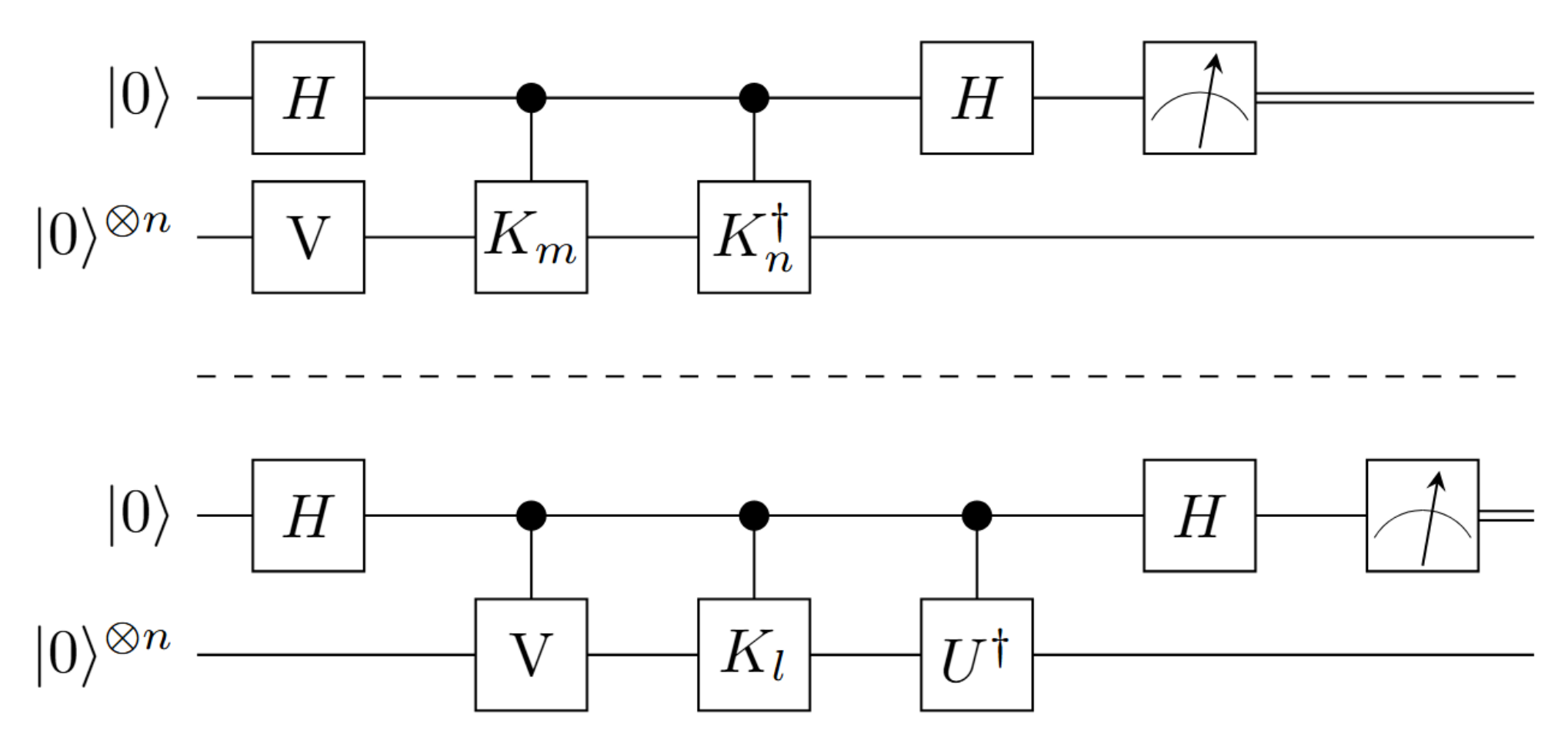}
    \caption{The Hadamard test to compute the components of the cost function. The top circuit outputs $\langle \bs{\psi}| \bs{\psi}\rangle$ in $L(L-1)/2$ runs. The bottom circuit outputs $\langle \bs{\psi}|f\rangle$ in $L$ runs.}
    \label{fig:VQLS_calcs}
\end{figure}

The implementation of the Hadamard test~\citep{cleve1998quantum} for the components of the cost function is shown in Figure~\ref{fig:VQLS_calcs}. The Hadamard test determines the expected real component of a unitary operation on a state (e.g., $\langle v| N_{nm} |v\rangle$) by measurement of an ancillary qubit. This is done by subtracting the probability of a zero state and one state on the ancilla, $P(0) - P(1)$. It is constructed by applying a Hadamard gate on the ancilla, controlling the unitary operation on the main register onto the ancilla, and applying another Hadamard gate on the ancilla. The uncontrolled $V(\bs{\theta})$ gate operating on the main register (Fig.~\ref{fig:VQLS_calcs}) prepares the state $|v\rangle$.

\vspace{-0.3cm}
\section{FEM Implementation based on VQLS}
\label{section:FEM_VQLS}
In this section, we describe the proposed approach to implement the finite element method (FEM) using VQLS. In the remainder of the manuscript, we refer to the proposed formulation as the quantum finite element method (Q-FEM). We first discuss the general methodology and apply it to a one-dimensional steady-state heat equation. 

Let $\mathcal{L}$ be a linear operator acting on an unknown field, $u$ that satisfies the expression $\mathcal{L} (u) = f$, where $f$ is a known field (i.e., forcing function). Considering the classical continuity and regularity requirements, the boundary and initial conditions, and using the FEM procedure results in a linear system of the form: $\mathbf{K} \mathbf{u} = \mathbf{f}$. The structure and properties of the stiffness matrix are dictated by the form of the linear operator, finite element discretization, shape functions used, node numbering, the method used for enforcing the boundary conditions, and in unstable problems, the stabilization terms. A critical consideration that differentiates quantum computer (QC) implementation of FEM (e.g., from the finite difference method) is that FEM allows for unstructured grids that facilitate the representation of complex geometries. The standard construction procedure for the global stiffness matrix and force vectors is the assembly operation:
\begin{equation}
    \mathbf{K} = \bigwedge_{e=1}^{n_{\mathrm{el}}} \mathbf{k}^{e},
\end{equation}
in which, $\mathbf{k}^{e}$ is the element stiffness matrix. The assembly operation performs mapping of the element degrees of freedom onto the global degrees of freedom. Element matrices associated with boundaries are treated in a slightly different manner. For simplicity, we ignore these complications in the current discussion and provide relevant details in the application of the approach to the heat equation.


An alternative construction procedure is based on the direct stiffness method~\citep{levy1953structural}. This approach consists of constructing `global' element stiffness matrices first, and summing the global element stiffness matrices to compute the stiffness matrix:
\begin{equation} \label{directStiffnessAssembly}
    \mathbf{K}=\displaystyle\sum_{e=1}^{n_{\mathrm{el}}} \mathbf{K}^e.
\end{equation}

The difference between the standard assembly and the direct stiffness assembly is that the latter delineates the local-to-global mapping from the addition operations. Each global element stiffness matrix, $\mathbf{K}^e$, has the same size as the stiffness matrix. Equation~\ref{directStiffnessAssembly} also implies that global element stiffness matrices are generated for each element. This condition will be relaxed using a concept of `unique element' as discussed in Section~\ref{sec:reduction_unitaries}. In this study, we employ the direct stiffness assembly approach, which allows a more straightforward QC implementation as mapping and summation operations are separated.  



Representing each global element matrix as a linear combination of unitary matrices, the global stiffness matrix is expressed as:
\begin{equation} \label{StiffnessUnitaryDecomposition}
    \mathbf{K} = \displaystyle\sum_{e=1}^{n_{\mathrm{el}}} \sum_{m = 1}^{M^e} c^e_m K^e_m.
\end{equation}
where $c^e_m$ and $K^e_m$ denote the weights and the unitary matrices, respectively. Noting that ${M^e}$ is a small number determined by the structure of the element stiffness matrix i.e.,~$2$ for linear elements or $4$ for quadratic elements, and that the number of elements in a finite element discretization scale with the number of degrees of freedom, the number of unitary matrices is of $\mathcal{O}(N)$. For fully or partially structured grids, this linear scaling could be broken and a constant set of unitaries could be sufficient. We note that, if the number of unitary matrices is reduced in the case of structured grids, the depth of the resulting circuits scales $\mathcal{O}(N)$. This effect is demonstrated in Section~\ref{sec:reduction_unitaries}.

Implementation of the QFEM approach therefore requires (1)~identifying the unitary matrices that efficiently represent the global element matrices (i.e., minimize $M^e$) and (2)~ developing the quantum circuits for each unitary matrix. The proposed construction procedure is outlined as follows: consider a library of circuits, $\mathcal{C}_I$, parameterized by an index set, $I:=\{i,j,\ldots\}$; this library is chosen such that it contains the corresponding circuit representation of each unitary matrix, $K^e_m$. We then define a generator function, $\mathcal{G}(e,m)$, that maps each unitary matrix with its corresponding circuit in the library. In the next section, we provide examples of the circuit construction in the context of the steady-state heat equation. 

In this study, the global force vector is constructed classically first and represented as a quantum state vector as discussed in Section \ref{sec:costfunction}.

\vspace{-0.3cm} \section{Application to the Steady-State Heat Equation} \label{section:Poisson_Eqn}
We first consider the 1-d steady-state heat equation with homogeneous Dirichlet boundary conditions. The strong form governing equation is given as follows:
\begin{equation} \label{strongFormEqn}
\begin{aligned}
    & c(x) \frac{d^2 u (x)}{d x^2} + b(x) = 0, \quad \forall x \in [x_1,x_2], \\
    & u(x_1) = 0, \quad u(x_2) = 0.
\end{aligned}
\end{equation}

For a given discretization of the domain, we show in the successive sections, how to generate the unitaries and the corresponding quantum circuits, that represent the stiffness matrix, for both linear and quadratic elements. The required changes in the unitaries and the quantum circuit to model the Neumann boundary condition and the non-zero Dirichlet boundary condition are discussed in a later subsection.

\vspace{-0.3cm} \subsection{Linear elements}
The domain is discretized with $n_{\mathrm{el}}$ linear elements, and the total number of nodes is $n_{\mathrm{el}} + 1$. For simplicity, the total number of internal nodes is chosen such that $N=2^{n}$, where $n$ is the number of qubits in the quantum circuit. The total number of nodes is $N+2$, accounting for the boundary nodes. Hence, for an $n$-qubit system, $n_{\mathrm{el}}$ is equal to $2^n +1$. The element numbering is selected such that $\mathbf{K}^0$ and $\mathbf{K}^{n_{\mathrm{el}}-1}$ are global-element matrices for boundary elements, and $\mathbf{K}^{e}$ for $e \in [1,n_{\mathrm{el}}-2]$ are internal element matrices.  

The global-element stiffness matrices for the boundary elements $\mathbf{K}^0$ and $\mathbf{K}^{n_{\mathrm{el}}-1}$, as well as for an internal element $\mathbf{K}^e$ are:
\begin{equation}
\begin{aligned}
  && \mathbf{K}^0 = \frac{c^0}{h^0}
\begin{bmatrix}
     1 & 0 & \cdots & 0 \\
     0 & 0 & \cdots & 0 \\
     \vdots & \vdots & \ddots & \vdots \\
     0 & 0 & \cdots & 0
\end{bmatrix}_{N \cross N},  \quad
\mathbf{K}^{n_{\mathrm{el}}-1} = \frac{c^{n_{\mathrm{el}}-1}}{h^{n_{\mathrm{el}}-1}}
\begin{bmatrix}
     0 & \cdots &  0 &  0 \\
     \vdots & \ddots & \vdots & \vdots \\
     0 & \cdots & 0 & 0 \\
     0 & \cdots & 0 & 1
\end{bmatrix}_{N \cross N}, \\
&& \mathbf{K}^e = \frac{c^e}{h^e}
\begin{bmatrix}
    \{0\}_{e-1 \cross e-1} & \{0\}_{e-1 \cross 1} & \cdots & \cdots & \cdots & \{0\}_{e-1 \cross 1} \\
   \{0\}_{1 \cross e-1} & 1 & -1 & 0 & \cdots & 0 \\
    \{0\}_{1 \cross e-1} & -1 & 1 & 0 & \cdots & 0 \\
    \{0\}_{1 \cross e-1} & 0 & 0 & 0 & \cdots & 0 \\
    \vdots & \vdots & \vdots &  \vdots & \ddots & \vdots \\
    \{0\}_{1 \cross e-1} & 0 & 0 & 0 & \cdots & 0
\end{bmatrix}_{N \cross N},
\end{aligned}
\end{equation}
where $h^{e}$ denotes element length, and $c^{e}$ is the diffusivity in element $e$. The global-element matrices are represented as a linear combination of unitary matrices as:
\begin{equation}
        \mathbf{K}^{e}= \displaystyle \frac{c^e}{h^e}
        \begin{cases}
            \displaystyle \frac{1}{2}\left( I - I_{e}^{-1} \right) & e\in\{0, n_{\mathrm{el}}-1\} \\
            \left( I - X_e \right) & e\in [1, n_{\mathrm{el}}-2]
    \end{cases}
\end{equation}
where $I$ is the identity matrix, $I_e^{-1}$ and $X_e$ are unitaries corresponding to the boundary and internal elements, respectively. One can obtain the linear combination of unitaries that represents the stiffness matrix as:
\begin{equation} \label{linearCaseUnitaryDecomposition}
    \mathbf{K} = \left( \frac{c^0}{ 2 h^0} + \sum_{e=1}^{n_{\mathrm{el}}-2} \frac{c^e}{h^e} + \frac{c^{n_{\mathrm{el}} - 1}}{2 h^{n_{\mathrm{el}} - 1}} \right) I - \frac{c^0}{ 2 h^0}  I_{0}^{-1}  -  \sum_{e=1}^{n_{\mathrm{el}}-2} \frac{c^e}{h^e} X_e - \frac{c^{n_{\mathrm{el}} - 1}}{2 h^{n_{\mathrm{el}} - 1}} I_{n_{\mathrm{el}} - 1}^{-1} .
\end{equation}

Equation~\eqref{linearCaseUnitaryDecomposition} indicates that the global stiffness matrix is represented using $n_{\mathrm{el}} + 1$ unitary matrices. Under certain conditions, it is possible to reduce the number of unitary matrices substantially, and these conditions are discussed in Section~\ref{sec:reduction_unitaries}. 

\subsubsection{Circuit generation} \label{circuitGenerationLinear}

We first proceed with the description of the most general case, where no unitary reduction is possible. The unitary matrices associated with the boundary elements and an internal element $e$ are:
\begin{equation} \label{linearElementUnitaries}
\begin{aligned}
 && I_0^{-1} = \frac{c^0}{2 h^0}
\begin{bmatrix}
    -1 & 0 & 0 & \cdots & 0 \\
    0 & 1 & 0 & \cdots & 0 \\
    0 & 0 & 1 &  & 0 \\
    \vdots & \vdots &  & \ddots &  \\
    0 & 0 & 0 &  & 1
\end{bmatrix}_{N \cross N}, \quad 
I_{n_{\mathrm{el}} - 1}^{-1} = \frac{c^{n_{\mathrm{el}} - 1}}{2 h^{n_{\mathrm{el}} - 1}}
\begin{bmatrix}
    1 & 0 & 0 & \cdots & 0 \\
    0 & 1 & 0 & \cdots & 0 \\
    0 & 0 & \ddots & \ddots & \vdots \\
    \vdots & \vdots & \ddots  & 1 &  0 \\
    0 & 0 & \cdots &  0 & -1
\end{bmatrix}_{N \cross N}, \\
&& X_e = \frac{c^e}{h^e}
\begin{bmatrix}
    \{I\}_{e-1 \cross e-1} & \{0\}_{e-1 \cross 1} & \cdots & \cdots & \cdots & \{0\}_{e-1 \cross 1} \\
   \{0\}_{1 \cross e-1} & 0 & 1 & 0 & \cdots & 0 \\
    \{0\}_{1 \cross e-1} & 1 & 0 & 0 & \cdots & 0 \\
    \{0\}_{1 \cross e-1} & 0 & 0 & 1 & \cdots & 0 \\
    \vdots & \vdots & \vdots &  \vdots & \ddots &  \\
    \{0\}_{1 \cross e-1} & 0 & 0 & 0 &  & 1
\end{bmatrix}_{N \cross N}.
\end{aligned}
\end{equation}

The quantum circuits for the unitaries of internal elements ${X}_e$ are generated based on three types of circuits, which are as follows:
\begin{itemize}
    \item Based on Pauli $X$ gates with control operations: $X_i^{0}$.
    \item Based on swap gates with control operations: $X_i^{1}$.
    \item Based on those generated by a series of controlled Pauli $X$ gates: $X_i^{2} \ldots X_i^{n-1}$.
\end{itemize}
The relationship between these circuit types and the internal element unitary, ${X}_e$, is established using the following generator function:
\begin{equation}\label{eq:Xcircuit_generator_function}
X_e = X_{2^{j}+ i \: 2^{j+1}} = X_{i}^{j}; \hspace{0.3cm} i \in \{0,1,\ldots,2^{n-j-1}-1 \}; \hspace{0.3cm} j \in \{0,1,\ldots,n-1\};
\end{equation}
where $e = 2^{j}+ i \: 2^{j+1}$ provides the mapping between the element and the circuit type. The index $i$ denotes the control operations on these circuits, while $j$ denotes the number of qubits $(j+1)$ that are flipped in the corresponding unitary matrix.

\begin{figure}[htb!]
    \centering
    \subfloat[][${X}_1 = X_0^{0}$]{
    \includegraphics[scale = 0.55]{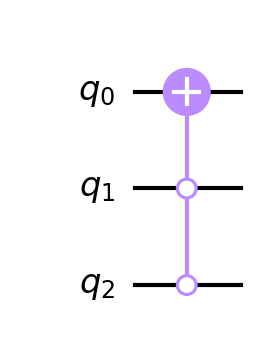}}
    \subfloat[][${X}_3 = X_1^{0}$]{
    \includegraphics[scale = 0.55]{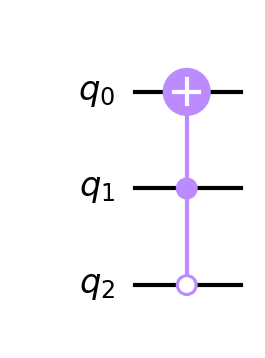}} 
    \subfloat[][${X}_5 = X_2^{0}$]{
    \includegraphics[scale = 0.55]{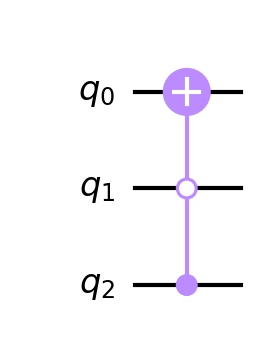}}
    \subfloat[][${X}_7 = X_3^{0}$]{
    \includegraphics[scale = 0.55]{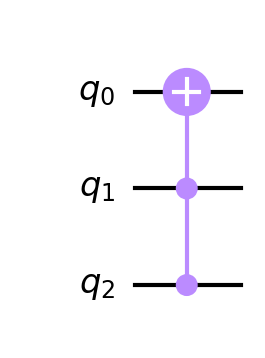}}
    \\
    \subfloat[][${X}_2 = X_0^{1}$]{
    \includegraphics[scale = 0.55]{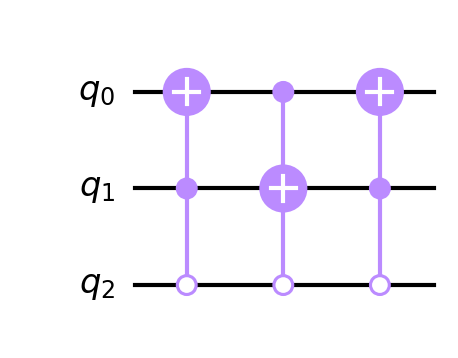}}
    \subfloat[][${X}_6 = X_1^{1}$]{
    \includegraphics[scale = 0.55]{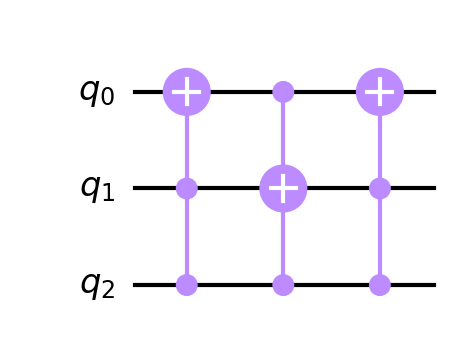}}
    \subfloat[][${X}_2 = X_0^{1}$]{
    \includegraphics[scale = 0.55]{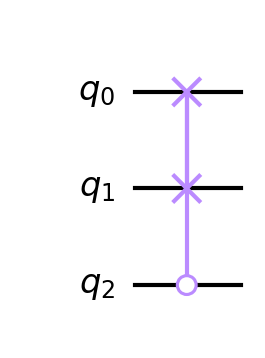}}
    \\
    \subfloat[][${X}_8 = X_0^{3}$]{ 
    \includegraphics[scale = 0.55]{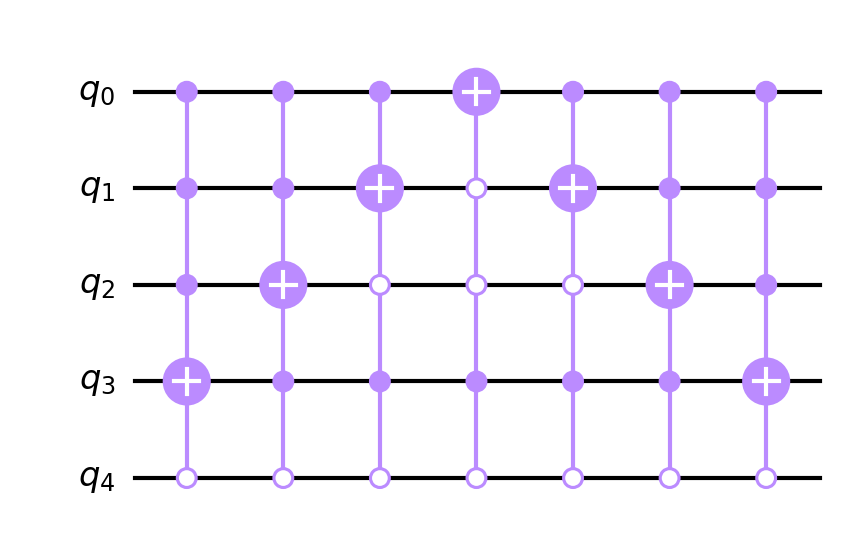}} 
    \subfloat[][${X}_{12} = X_1^{2}$]{ 
    \includegraphics[scale = 0.55]{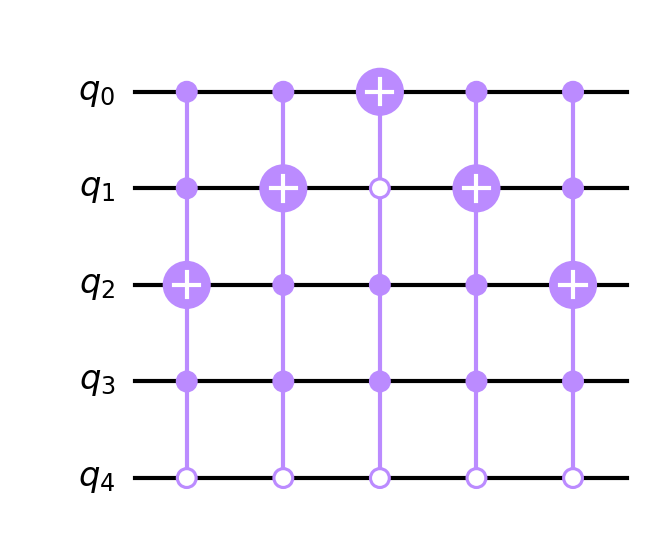}} 
    \caption{Examples of quantum circuits for unitaries corresponding to internal elements for a 3-qubit system (a)-(g) and 5-qubit system (h)-(i).}
    \label{fig:3-qubit-internal-cicuits}
\end{figure}

The three circuit types for an $n$ qubit system are obtained as follows:
\begin{enumerate}
    \item \textbf{$X^0_{i}$ circuit}: one of the qubits is flipped in the circuit corresponding to this unitary matrix. This is achieved by applying the Pauli $X$ gate to the $q_0$ (first) qubit controlled by the remaining qubits. The subscript $i$ denotes a particular control operation out of $2^{n-1}$ possibilities on the remaining $n-1$ qubits, as each control qubit can be either open-controlled or closed-controlled. In particular, $i$ denotes the control operation corresponding to the $(i+1)^{\mathrm{th}}$ basis state for the control qubits. This is illustrated in Fig.~\ref{fig:3-qubit-internal-cicuits}a-d for the 3-qubit example. For $n=3$, $i=0$ denotes open control operations on the control qubits based on the first basis state $\ket{00}$, while $i=1$ denotes open and closed control operations on $q_2$ and $q_1$ qubits, respectively, based on the second basis state $|01 \rangle$.
    
    \item \textbf{$X^1_{i}$ circuit}: the number of qubits flipped in the circuit corresponding to this unitary matrix is $2$. In this circuit, a swap gate is applied between the first two qubits $q_0$ and $q_1$, and $i$ denotes a particular control operation out of $2^{n-2}$ possibilities on the remaining $n-2$ qubits. As deduced for $X^0_{i}$ circuit, the control operation $i$ corresponds to the $(i+1)^{\mathrm{th}}$ basis state for the remaining $n-2$ qubits. Corresponding circuits for the 3-qubit case are shown in Fig.~\ref{fig:3-qubit-internal-cicuits}e-f. As an example, for $n=3$ case, $i=0$ denotes open control operation on the control qubits based on the first basis state $\ket{0}$, while $i=1$ denotes closed control operation on $q_3$ based on the second basis state $|1 \rangle$. Naturally, the three layers of operations could be replaced by a single controlled swap gate as well, shown in Fig.~\ref{fig:3-qubit-internal-cicuits}g for $X^1_0$ circuit. 
    
    \item \textbf{$X_i^{j}$ circuits} with $j \in \{ 2, 3, \cdots, n-1\}$: the number of qubits flipped in this circuit corresponding to the unitary matrix is $j+1$. The depth of the circuit is $2j + 1$, and each layer has controlled Pauli $X$ gates. The number of pure control qubits (where no $X$ gate is applied) is $n-j-1$, which are controlled based on their $(i+1)^{\mathrm{th}}$ basis state. The overall strategy is to design a series of controlled $X$ circuits to generate the circuit corresponding to the transformation in the unitary matrix, when the number of qubits flipped is more than 2. For example, a basis state $|{\color{red}1\cdots0}0111 \cdots 1 \rangle$  transforms to $|{\color{red}1\cdots0}1000 \cdots 0 \rangle$ state and vice-versa, where the qubits highlighted in red are pure control qubits and the qubits in black are the ones that are flipped in this circuit. The first $j+1$ controlled $X$ gates in the series of gates flip a qubit starting from the most dominant to the least, excluding the list of pure control qubits. This transforms the $|{\color{red}1\cdots0}0111 \cdots 1 \rangle$ basis state to $|{\color{red}1\cdots0}1000 \cdots 0 \rangle$ state, and $|{\color{red}1\cdots0}1000 \cdots 0 \rangle$ state to $|{\color{red}1\cdots0}1000 \cdots 1 \rangle$ state. To correct the latter, the remaining $j$ controlled $X$ gates are applied such that $|{\color{red}1\cdots0}1000 \cdots 1 \rangle$ state transforms to $|{\color{red}1\cdots0}0111 \cdots 1 \rangle$, with qubits being flipped starting from $q_1$ to the most dominant qubit, excluding pure control qubits. This procedure is demonstrated for the following two examples.
    
    First, given a $5$-qubit system, a series of operations for $X^3_0$ circuit (shown in Fig.~\ref{fig:3-qubit-internal-cicuits}h) are listed: firstly a $|{\color{red}0}0111 \rangle$ state is transformed to $|{\color{red}0}1111 \rangle$, then from $|{\color{red}0}1111 \rangle$ to $|{\color{red}0}1011 \rangle$, and then from $|{\color{red}0}1011 \rangle$ to $|{\color{red}0}1001 \rangle$ and finally from $|{\color{red}0}1001 \rangle$ to $|{\color{red}0}1000 \rangle$. Then reverse transformations in the opposite order are done starting from $|{\color{red}0}1001 \rangle$ (leaving the penultimate one) to $|{\color{red}0}1011 \rangle$ till the first state ($|{\color{red}0}0111 \rangle$) to begin with is obtained. These operations are done using controlled Pauli $X$ gates, with the control operations in non-pure control qubits changing accordingly.

    Second, for a $5$-qubit system and $X^2_1$ circuit shown in Fig.~\ref{fig:3-qubit-internal-cicuits}i, these operations are listed: firstly a $|{\color{red}01}011 \rangle$ state is transformed to $|{\color{red}01}111 \rangle$, then from $|{\color{red}01}111 \rangle$ to $|{\color{red}01}101 \rangle$, and finally from $|{\color{red}01}101 \rangle$ to $|{\color{red}01}100 \rangle$. Then reverse transformations in the opposite order are performed starting from $|{\color{red}01}101 \rangle$ state (leaving the penultimate one) to $|{\color{red}01}111 \rangle$ state till the first state ($|{\color{red}01}011 \rangle$) to begin with is obtained. All of these operations are performed using controlled Pauli $X$ gates.
\end{enumerate}

The quantum circuits for the unitaries corresponding to boundary elements are explained below:
\begin{enumerate}
    \item The $I_0^{-1}$ circuit is a controlled $Ry$ rotation gate with $\theta = 2 \pi$ as a parameter on $q_{0}$ qubit and open control operations on remaining $n-1$ qubits. This is followed by a controlled $Z$ gate with closed control on the $q_0$ qubit and open controls on the remaining $n-1$ qubits. 
    \item The $I_{n_{\mathrm{el}} - 1}^{-1}$ circuit is a controlled $Z$ gate with closed control operations on the $n$ qubits.
\end{enumerate}

\begin{figure}[htb!]
    \centering
    \subfloat[][${I}_0^{-1}$]{
    \includegraphics[scale = 0.65]{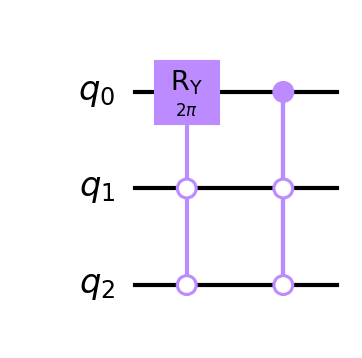}}
    \subfloat[][${I}_8^{-1}$]{
    \includegraphics[scale = 0.65]{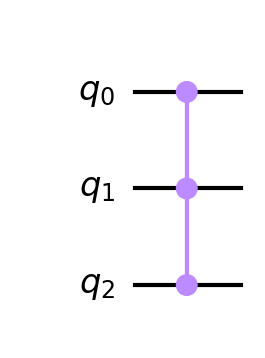}}
    \caption{Examples of quantum circuits for unitaries corresponding to boundary elements for a 3-qubit system.}
    \label{fig:3-qubit-boundary-cicuits}
\end{figure}

The quantum circuits for the unitaries corresponding to boundary elements are shown in Fig.~\ref{fig:3-qubit-boundary-cicuits} for a 3-qubit system.

\vspace{-0.3cm} \subsection{Quadratic elements}

\begin{figure}[htb!]
    \centering
    \includegraphics[scale = 1.0]{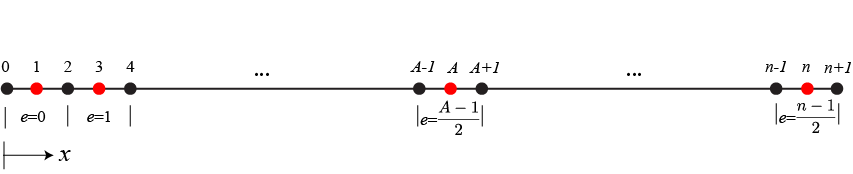}
    \caption{The discretization of the domain into quadratic elements.}
    \label{fig:problem_quadratic_element}
\end{figure}

Figure \ref{fig:problem_quadratic_element} shows the discretization of the problem domain using quadratic elements. Considering an $n$ qubit register, the total number of internal nodes is $\Tilde{N} = 2^n - 1$. The number of elements is $n_{\mathrm{el}} = (\Tilde{N}+1)/2 = 2^{n-1}$. Let us denote the position coordinate of node $A$ as $x_A$, and the internal nodes in the domain are such that $1 \leq A \leq \Tilde{N}$, and the position coordinates of boundary nodes are $x_0$ and $x_{\Tilde{N}+1}$. The length of an element is given by $h_e = x_{2e+2} - x_{2e}$. It is worthy of note that the size of a state vector that can be represented in the quantum computer is $N = 2^n = \Tilde{N}+1$. To accommodate this, an auxiliary degree of freedom is added to the system, and hence the size of the global stiffness matrix and global force vector are $N \times N$ and $N$, respectively.

As per the discretization of the problem domain shown in Fig.~\ref{fig:problem_quadratic_element}, the boundary elements correspond to $e=0$ and $e={n_{\mathrm{el}}-1}$, while the internal elements are $e \in [1,n_{\mathrm{el}}-2]$. For an $n$-qubit system, the global-element matrices for boundary elements $\mathbf{K}^0$ and $\mathbf{K}^{n_{\mathrm{el}}-1}$ are:
\begin{equation}
\begin{aligned}
  &&  \mathbf{K}^0 = \frac{c^0}{3 h^0}
\begin{bmatrix}
   16 & -8 & 0 & 0 & \cdots & 0 \\
    -8 & 7 & 0 & 0 &\cdots & 0 \\
    0 & 0 & 0 & 0 & \cdots & 0 \\ 
    0 & 0 & 0 & 0 & \cdots & 0 \\ 
    \vdots & \vdots & \vdots & \vdots & \ddots & \vdots \\
    0 & 0 & 0 & 0 & \cdots & 0
\end{bmatrix},
\mathbf{K}^{n_{\mathrm{el}}-1} = \frac{c^{n_{\mathrm{el}}-1}}{3 h^{n_{\mathrm{el}}-1}} \:
\begin{bmatrix}
    0 & \cdots & 0 & 0 & 0 & 0 \\
    \vdots & \ddots & \vdots & \vdots & \vdots & 0 \\
    0 & \cdots & 0 & 0 & 0 & 0 \\ 
    0 & \cdots & 0 & 0 & 0 & 0 \\ 
    0 & \cdots & 0 & 0 & 7 & -8 \\
    0 & \cdots & 0 & 0 & -8 & 16
\end{bmatrix},
\end{aligned}
\end{equation}
and for an arbitrary internal element $\mathbf{K}^e$ is:
\begin{equation}
\mathbf{K}^e = \frac{c^e}{3 h^e}
\begin{bmatrix}
    \{0\}_{2e-1 \times 2e-1} & \{0\}_{2e-1 \times 1} & \cdots & \cdots & \cdots & \cdots & \{0\}_{2e-1 \times 1} \\ 
   \{0\}_{1 \times 2e-1} & 7 & -8 & 1 & 0 & \cdots & 0 \\
    \{0\}_{1 \times 2e-1} & -8 & 16 & -8 & 0 &\cdots & 0 \\
    \{0\}_{1 \times 2e-1} & 1 & -8 & 7 & 0 & \cdots & 0 \\ 
    \{0\}_{1 \times 2e-1} & 0 & 0 & 0 & 0 & \cdots & 0 \\ 
    \vdots & \vdots & \vdots  & \vdots & \vdots & \ddots & \vdots \\
    \{0\}_{1 \times 2e-1} & 0 & 0 & 0 & 0 & \cdots & 0
\end{bmatrix}_{N \cross N }.
\end{equation}

The global-element matrices for quadratic elements are expressed as a linear combination of unitary matrices as follows
\begin{equation} \label{element_unitaries_decomposition}
        \mathbf{K}^{e}= \displaystyle \frac{c^e}{3h^e}
        \begin{cases}
            \displaystyle \frac{1}{2}\left( 15 \: {I} + 9 \: {Z}_{e} - 16 \: {X}_{ae+1} - 8 \: {I}^{-1}_{ae+1} \right) & e\in\{0, n_{\mathrm{el}}-1\} \\
            15 \: {I} - 8 \: {X}_{2e} - 8 \: {X}_{2e+1} + \Tilde{{X}}_{2e} & e\in [1, n_{\mathrm{el}}-2]
    \end{cases},
\end{equation}
%
where ${X}_{(\cdot)}$ is the same unitary matrix representation used for linear elements as described in Eq.~\eqref{linearElementUnitaries}, except with $a=(2n_{\mathrm{el}}-3)/(n_{\mathrm{el}}-1)$. The explicit expression of the unitary matrix $\Tilde{{X}}_{2e}$ illustrates the structure of the additional matrices that need to be constructed for the internal elements:
\begin{equation}
    \Tilde{{X}}_{2e} = 
\begin{bmatrix}
    \{I\}_{2e-1 \times 2e-1} & \{0\}_{2e-1 \times 1} & \cdots & \cdots & \cdots & \cdots & \{0\}_{2e-1 \times 1} \\ 
   \{0\}_{1 \times 2e-1} & 0 & 0 & 1 & 0 & \cdots & 0 \\
    \{0\}_{1 \times 2e-1} & 0 & 1 & 0 & 0 &\cdots & 0 \\
    \{0\}_{1 \times 2e-1} & 1 & 0 & 0 & 0 & \cdots & 0 \\ 
    \{0\}_{1 \times 2e-1} & 0 & 0 & 0 & 1 & \cdots & 0 \\ 
    \vdots & \vdots & \vdots  & \vdots & \vdots & \ddots &  \\
    \{0\}_{1 \times 2e-1} & 0 & 0 & 0 & 0 &  & 1
\end{bmatrix}_{N \cross N }.
\end{equation}
%
For $e=0$, the explicit expression of the unitary matrices ${Z}_{e}$ and ${I}^{-1}_{ae+1}$ have the form:
\begin{equation}
{Z}_0 = 
\begin{bmatrix}
   1 & 0 & 0 & 0 & \cdots & 0 \\
    0 & -1 & 0 & 0 &\cdots & 0 \\
    0 & 0 & 1 & 0 & \cdots & 0 \\ 
    0 & 0 & 0 & 1 &  & 0 \\ 
    \vdots & \vdots & \vdots &  & \ddots &  \\
    0 & 0 & 0 & 0 &  & 1
\end{bmatrix}_{N  \cross N}, \quad
{I}^{-1}_{1} = 
\begin{bmatrix}
   -1 & 0 & 0 & 0 & \cdots & 0 \\
    0 & -1 & 0 & 0 &\cdots & 0 \\
    0 & 0 & 1 & 0 & \cdots & 0 \\ 
    0 & 0 & 0 & 1 &  & 0 \\ 
    \vdots & \vdots & \vdots &  & \ddots &  \\
    0 & 0 & 0 & 0 &  & 1
\end{bmatrix}_{N  \cross N}.
\end{equation}

Using Eq.~\eqref{element_unitaries_decomposition}, the global stiffness matrix is expressed as a linear combination of the unitary matrices as:

\begin{multline} \label{eq:quadratic_elements_all_unitaries}
    \mathbf{K} = \left( \frac{5 c^0}{2 h^0} + \displaystyle\sum_{e=1}^{n_{\mathrm{el}}-2} \frac{15 c^e}{3 h^e} + \frac{5 c^{n_{\mathrm{el}}-1}}{2 h^{n_{\mathrm{el}}-1}} \right) \: I + \frac{c^0}{h^0} \left( \frac{3}{2} \: Z_{0} - \frac{8}{3} \: X_{1} - \frac{4}{3} \: I_{1}^{-1} \right) \\
    + \displaystyle\sum_{e=1}^{n_{\mathrm{el}}-2} \left( \frac{c^e}{h^e} \left( - \frac{8}{3} \: X_{2e} - \frac{8}{3} \: X_{2e+1} + \frac{1}{3} \: \Tilde{X}_{2e} \right) \right) \\
    + \frac{c^{n_{\mathrm{el}}-1}}{h^{n_{\mathrm{el}}-1}} \left( \frac{3}{2} \: Z_{n_{\mathrm{el}}-1} - \frac{8}{3} \: X_{2(n_{\mathrm{el}}-1)} - \frac{4}{3} \: I_{2(n_{\mathrm{el}}-1)}^{-1} \right).
\end{multline}

It suffices to develop the corresponding circuits for ${Z}_{(\cdot)}$ and $\Tilde{{X}}_{(\cdot)}$ to fully describe the stiffness matrix. In general, for $n_{\mathrm{el}}$ quadratic elements, one would need $3 n_{\mathrm{el}} + 1$ unitaries to represent the global stiffness matrix if each quadratic element has a different length or a different diffusivity.

\subsubsection{Circuit generation}
We generate quantum circuits corresponding to the unitaries whose linear combination forms the global matrix for quadratic elements as shown in Eq.~\eqref{eq:quadratic_elements_all_unitaries}. First, we develop generator functions for unitary matrices corresponding to internal elements. Generation of circuits corresponding to ${X}_{e}$ are described in Section~\ref{circuitGenerationLinear}, as they are identical to the linear elements and are generated using the functions described in Eq.~\eqref{eq:Xcircuit_generator_function}. 

The $\Tilde{{X}}_{(\cdot)}$ circuits are generated using two types of circuits, which are:
\begin{itemize}
    \item Based on Pauli $X$ gates with control operations: $\Tilde{X}_i^{1}$,
    \item Based on those generated by a series of controlled Pauli $X$ gates: $\Tilde{X}_i^{2} \ldots \Tilde{X}_i^{n-1}$,
\end{itemize}
where $i$ denotes the control operations on these circuits, while $j$ indicates the number of qubits flipped to generate the corresponding unitary matrix. We use the following generator function to assign $\Tilde{{X}}_{(\cdot)}$ from the two types of circuits: 
\begin{equation}\label{eq:Xtilde_generator_function}
\Tilde{{X}}_A = \Tilde{{X}}_{2^{j}+i \: 2^{j+1}} = \Tilde{X}_{i}^{j}; \hspace{0.3cm} i \in \{0,1,\ldots,2^{n-j-1}-1 \}; \hspace{0.3cm} j \in \{1,\ldots,n-1\},    
\end{equation}
with $A = 2^{j}+ i \: 2^{j+1}$ and $A=2e$ providing the mapping between the element and the circuit type. We explain below the generation of these circuits for a $n$ qubit system.
\begin{enumerate}
    \item \textbf{$\Tilde{X}^{1}_{i}$ circuit}: one of the qubits is flipped in this circuit to obtain the corresponding unitary matrix. This is achieved by applying the Pauli $X$ gate to the $q_1$ qubit, which is controlled by the remaining qubits. The $q_0$ qubit is always close-controlled, and $i$ denotes the control operations on the other qubits. In particular, $i$ denotes the control operation corresponding to the $(i+1)^{th}$ basis state for the remaining $n-2$ qubits.
    \item \textbf{$\Tilde{X}^{j}_{i}$ circuits} with $j \in \{2,3, \cdots, n-1 \}$: the number of qubits that are flipped in this circuit to obtain the corresponding unitary matrix is $j$. The depth of the circuit is $2j-1$, and each layer has controlled Pauli $X$ gates. The $q_0$ qubit is always close-controlled, and the number of pure control qubits (where no $X$ gate is applied) is $n-j-1$. These are controlled based on the $(i+1)^{\mathrm{th}}$ basis state. The overall strategy is to design a series of controlled $X$ circuits to generate the circuit corresponding to the transformation in the unitary matrix, and the strategy is similar to the one used for $X_{i}^{j}$ circuits. For example, a $|{\color{red}1 \cdots 0}01 \cdots 11 \rangle$ state is transformed to a $|{\color{red}1 \cdots 0}10 \cdots 0 1 \rangle$ state and vice-versa, where qubits highlighted in red are the pure control qubits and those in black are the qubits that are flipped in the circuit, except for the $q_0$ qubit. 
    
    A series of controlled $X$ circuits to obtain the $\Tilde{X}^{2}_{0}$ circuit and the corresponding unitary matrix for a 3-qubit system is demonstrated. In this example, there are no pure control qubits. First, a $|011\rangle$ state is transformed to $|111\rangle$ state, then $|111\rangle$ is transformed to $|101 \rangle$. Then, reverse transformations are performed in the opposite order, starting from the $|111\rangle$ state (leaving the penultimate one) to the $|011\rangle$ state (i.e.,~the final state).
\end{enumerate}

\begin{figure}[htb!]
    \centering
    \subfloat[][$\Tilde{{X}}_2 = \Tilde{X}^{1}_{0}$]{
    \includegraphics[scale = 0.55]{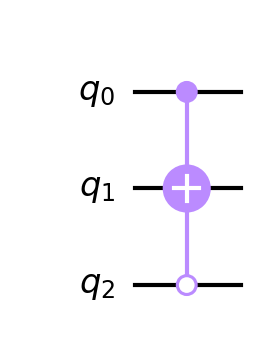}}
    \subfloat[][$\Tilde{{X}}_4 = \Tilde{X}^{2}_{0}$]{
    \includegraphics[scale = 0.55]{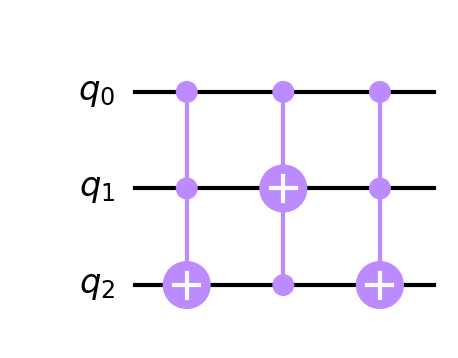}}
    \caption{The quantum circuits for $\Tilde{{X}}_{(\cdot)}$ unitaries corresponding to internal elements of a quadratic interpolation for a 3-qubit system.}
    \label{fig:3-qubit-Xitlde-cicuits}
\end{figure}

Figure \ref{fig:3-qubit-Xitlde-cicuits} shows the quantum circuits for the $\Tilde{{X}}_{(\cdot)}$ unitaries corresponding to internal elements for a 3-qubit system, and these are obtained using the generator function given in Eq.~\eqref{eq:Xtilde_generator_function}.  

\begin{figure}[htb!]
    \centering
    \subfloat[][${Z}_0$]{
    \includegraphics[scale = 0.55]{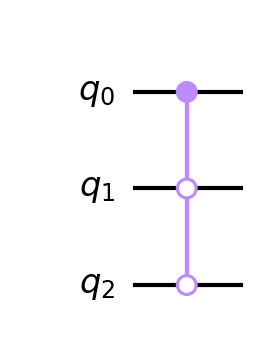}}
    \subfloat[][${Z}_{(n_{\mathrm{el}} - 1)}$]{
    \includegraphics[scale = 0.55]{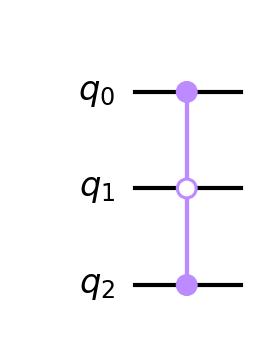}}
     \subfloat[][${I}_{1}^{-1}$]{
    \includegraphics[scale = 0.55]{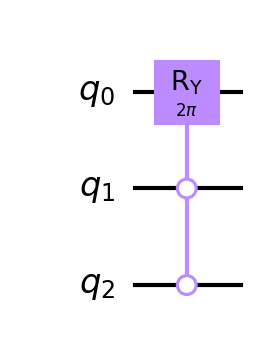}}
    \subfloat[][${I}^{-1}_{2(n_{\mathrm{el}} - 1)}$]{
    \includegraphics[scale = 0.55]{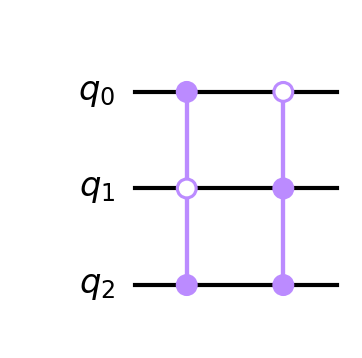}}
    \caption{The quantum circuits for unitaries corresponding to boundary elements of a quadratic interpolation for a 3-qubit system.}
    \label{fig:3-qubit-quadratic-boundary-cicuits}
\end{figure}

The quantum circuits for the $Z_{(\cdot)}$ unitaries corresponding to boundary elements are explained below:
\begin{itemize}
    \item The $Z_{0}$ circuit is a controlled $Z$ gate with closed control on the $q_0$ qubit and open control operations on remaining $n-1$ qubits.
    \item The $Z_{(n_{\mathrm{el}}-1)}$ circuit is a controlled $Z$ gate with open control on $q_1$ and closed control operations on the remaining $n-1$ qubits.
\end{itemize}
The boundary quadratic elements has additional $I_{(\cdot)}^{-1}$ circuits which are explained below:
\begin{itemize}
    \item The $I_{1}^{-1}$ circuit is obtained by a controlled $Ry$ rotation gate with $\theta = 2 \pi$ as a parameter on $q_0$ qubit, with open control operation on the remaining $n-1$ qubits. 
    \item The $I_{2(n_{\mathrm{el}} - 1)}^{-1}$ circuit is obtained by two controlled $Z$ gates. In the first $Z$ gate, $q_0$ qubit is close-controlled, $q_1$ is open-controlled, and the remaining $n-2$ qubits are close-controlled. For the second $Z$ gate, $q_0$ qubit is open-controlled, $q_1$ is close-controlled, and the remaining $n-2$ qubits are close-controlled. 
\end{itemize}
The circuits for unitaries corresponding to the boundary elements for a 3-qubit system are shown in Fig.~\ref{fig:3-qubit-quadratic-boundary-cicuits}, except for ${X}_1$ and ${X}_{6}$, which are already shown in Fig.~\ref{fig:3-qubit-internal-cicuits}.

As discussed earlier, the actual size of the vector that can be represented in the quantum computer is $N=\Tilde{N}+1$. To account for the additional degree of freedom, the size of the global stiffness matrix is adjusted, and the additional unitaries (and the corresponding circuits) are accounted for in the VQLS algorithm. The global stiffness matrix has an additional row and column as shown below:
\begin{equation}
\mathbf{K} = 
\begin{bmatrix}
   \mathbf{K}_{\Tilde{N} \times \Tilde{N}}  & \{0 \}_{\Tilde{N}  \cross 1} \\
    \{0\}_{1 \cross \Tilde{N}} & 1
\end{bmatrix}_{N \cross N}.
\end{equation}
The additional row and column can be written as a combination of unitaries, as shown below:
\begin{equation}
\frac{1}{2} \left( {I} - {I}^{-1}_{2 n_{\mathrm{el}}} \right), 
\end{equation}
where 
\begin{equation}
{I}^{-1}_{2n_{\mathrm{el}}} = 
\begin{bmatrix}
   1 & 0 & 0 & 0 & \cdots & 0 \\
    0 & 1 & 0 & 0 &\cdots & 0 \\
    0 & 0 & 1 & 0 & \cdots & 0 \\ 
    \vdots & \vdots & \ddots & \ddots & \ddots &  \vdots \\
    0 & 0 &  & 0 & 1 & 0 \\ 
    0 & 0 & 0 & \cdots & 0 & -1
\end{bmatrix}_{N \cross N }.    
\end{equation}
The $N^{\mathrm{th}}$ entry in the force vector can be arbitrarily chosen, before normalization, but for simplicity, we take it as zero for our simulations.

\subsection{Concatenation of global element matrices} \label{sec:reduction_unitaries}
In this section, we present a procedure for reducing the number of global element stiffness matrices by defining the concept of identical non-interacting elements. Two elements that are identical non-interacting have the same element stiffness matrix and do not interact with each other. In the context of the 1-d heat equation, global element matrices for identical non-interacting elements, e.g. non-adjacent elements with identical element lengths and diffusion coefficients, are combined. These elements form an identical non-interacting element group, and the total number of identical non-interacting elements is determined accordingly for a given problem and discretization. The global stiffness matrix is then expressed as the sum of global element matrices corresponding only to these identical non-interacting elements. The concatenated matrices for each unique element are decomposed into unitaries, which are represented by a series of quantum circuits corresponding to the unitary matrices (other than the identity matrix) of the individual elements within each identical non-interacting element group. The following sections illustrate this process for both linear and quadratic element discretizations.

\subsubsection{Linear elements}
The number of identical non-interacting elements is denoted as $n^{\mathrm{idn}}_{\mathrm{el}}$. For generality, it is assumed that the boundary elements are treated as identical non-interacting elements, although, this assumption can easily be relaxed in special cases. By applying the concept of identical non-interacting elements, the global stiffness matrix can be expressed as follows:
\begin{equation} 
    \mathbf{K} = \left( \frac{c^0}{ 2 h^0} + \sum_{e=1}^{n^{\mathrm{idn}}_{\mathrm{el}}-2} \frac{c^e}{h^e} + \frac{c^{n^{\mathrm{idn}}_{\mathrm{el}}-1}}{2 h^{n^{\mathrm{idn}}_{\mathrm{el}}-1}} \right) I - \frac{c^0}{ 2 h^0}  I_{0}^{-1}  -  \sum_{e=1}^{n^{\mathrm{idn}}_{\mathrm{el}}-2} \frac{c^e}{h^e} X^{\mathrm{idn}}_e - \frac{c^{n^{\mathrm{idn}}_{\mathrm{el}}-1}}{2 h^{n^{\mathrm{idn}}_{\mathrm{el}}-1}} I_{n_{\mathrm{el}}-1}^{-1},
\end{equation}
where $X^{\mathrm{idn}}_e$ denotes the unitary matrix for the $e^{\mathrm{th}}$ identical non-interacting element. The expression for $X^{\mathrm{idn}}_e$ unitary is given as:
\begin{equation}
    X^{\mathrm{idn}}_e = X_{e_0} X_{e_1} \cdots X_{e_{m}},
\end{equation}
where $e_{(\cdot)}$ denotes the element number of the non-adjacent elements in the original discretization that are being concatenated, and $X_{(\cdot)}$ denotes the corresponding unitary for that element. The concatenated unitary is formed by applying the individual unitaries of the elements within the identical non-interacting element group, and the circuit depth of the concatenated unitary is the sum of circuit depths of unitaries within the group. In the special case, when all elements have identical lengths and diffusion coefficients, the expression for $X^{\mathrm{idn}}_{1}$ and $X^{\mathrm{idn}}_{2}$ are as shown below:
\begin{eqnarray}
    X^{\mathrm{idn}}_{1} &=& X_{1} X_{3} \cdots X_{n_{\mathrm{el}}-2},\\
    X^{\mathrm{idn}}_{2} &=& X_{2} X_{4} \cdots X_{n_{\mathrm{el}}-3}.
\end{eqnarray}

If the boundary elements are combined similarly, then the number of unitaries required to represent the global stiffness matrix is $4$. 

As an example, unitary concatenation is demonstrated below for a 2-qubit system with linear elements. The number of internal nodes and internal elements is $4$ and $3$, respectively, with Dirichlet boundary conditions imposed on the boundary nodes. In this case, there are only $3$ internal element unitaries, i.e., $X_1$, $X_2$, and $X_3$, of which $X_1$ and $X_3$ are non-interacting and can be concatenated if made of identical lengths and diffusivity coefficients. The $X_1$, $X_3$, and concatenated $X^{\mathrm{idn}}_1 = X_1 X_3$ unitaries are shown below:

\begin{equation}
\begin{aligned}
  && X_1 =
\begin{bmatrix}
     0 & 1 & 0 & 0 \\
     1 & 0 & 0 & 0 \\
     0 & 0 & 1 & 0 \\
     0 & 0 & 0 & 1
\end{bmatrix}, \quad
 X_3 =
\begin{bmatrix}
     1 & 0 & 0 & 0 \\
     0 & 1 & 0 & 0 \\
     0 & 0 & 0 & 1 \\
     0 & 0 & 1 & 0
\end{bmatrix}, \quad 
X^{\mathrm{idn}}_1 =
\begin{bmatrix}
     0 & 1 & 0 & 0 \\
     1 & 0 & 0 & 0 \\
     0 & 0 & 0 & 1 \\
     0 & 0 & 1 & 0
\end{bmatrix}.
\end{aligned}
\end{equation}

\begin{figure}[htb!]
    \centering
    \subfloat[][]{
    \includegraphics[scale=0.6]{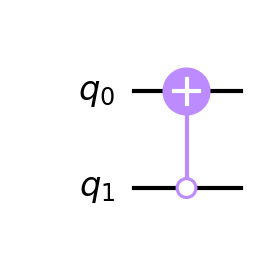}}
    \subfloat[][]{
    \includegraphics[scale=0.6]{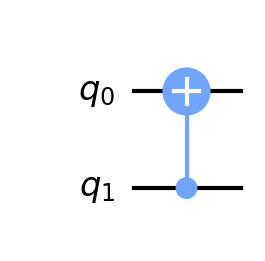}}
    \subfloat[][]{
    \includegraphics[scale=0.6]{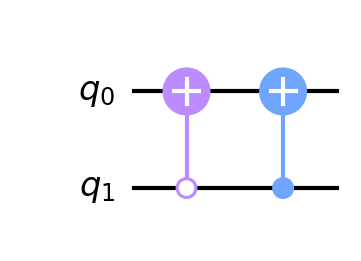}}
    \caption{Internal element circuits for 2-qubit system with linear elements: (a) $X_1$, (b) $X_3$, and concatenated circuit in (c) $X^\mathrm{idn}_{1} = X_1 X_3$.}
    \label{fig:concatenated_circuit_linear_2_qubit}
\end{figure}

Figure~\mbox{\ref{fig:concatenated_circuit_linear_2_qubit}} shows the internal element circuits for linear elements without concatenation ($X_1$ and $X_3$), and with concatenation $X^\mathrm{idn}_{1}$. The circuit depth of the concatenated unitary is $2$, and is obtained by applying $X_1$ and $X_3$ as evident in Fig.~\mbox{\ref{fig:concatenated_circuit_linear_2_qubit}}c.

\begin{figure}[htb!]
    \centering
    \subfloat[][]{
    \includegraphics[scale=0.4]{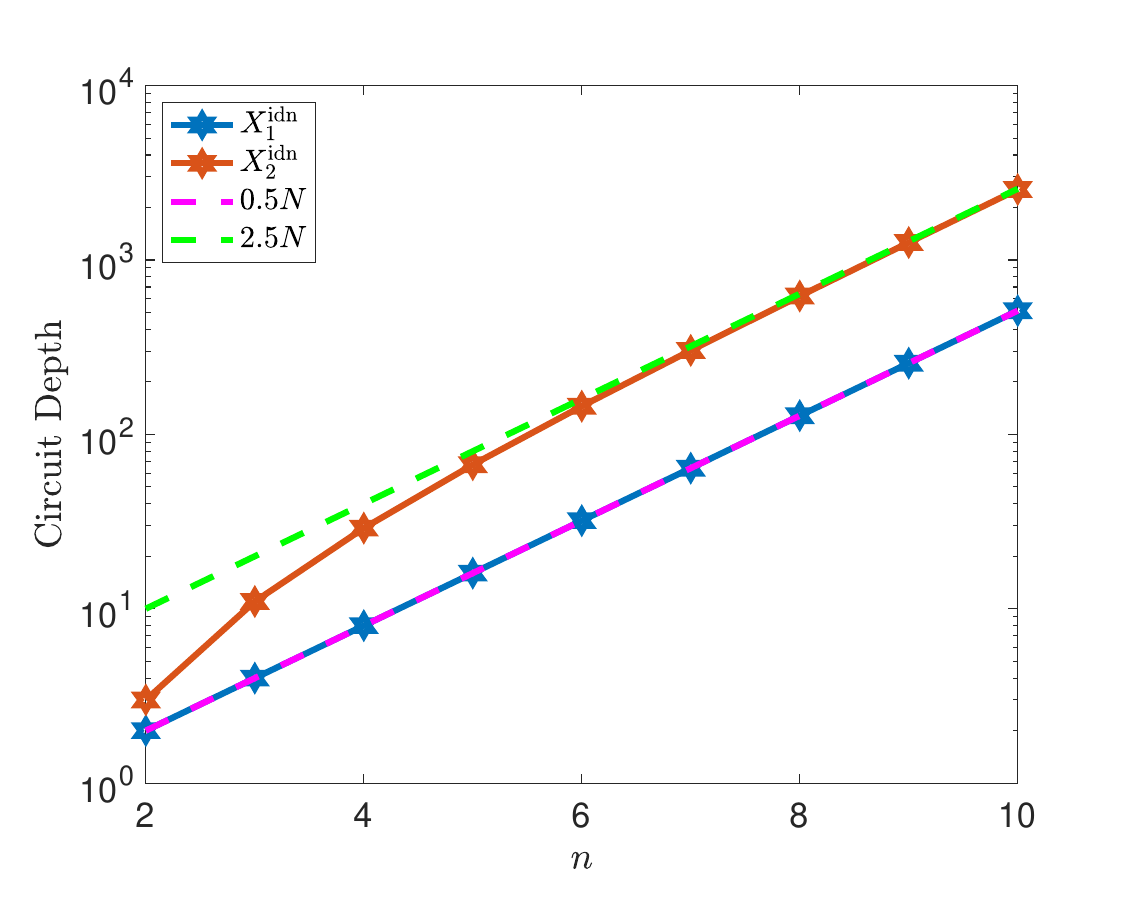}}
    \subfloat[][]{
    \includegraphics[scale=0.4]{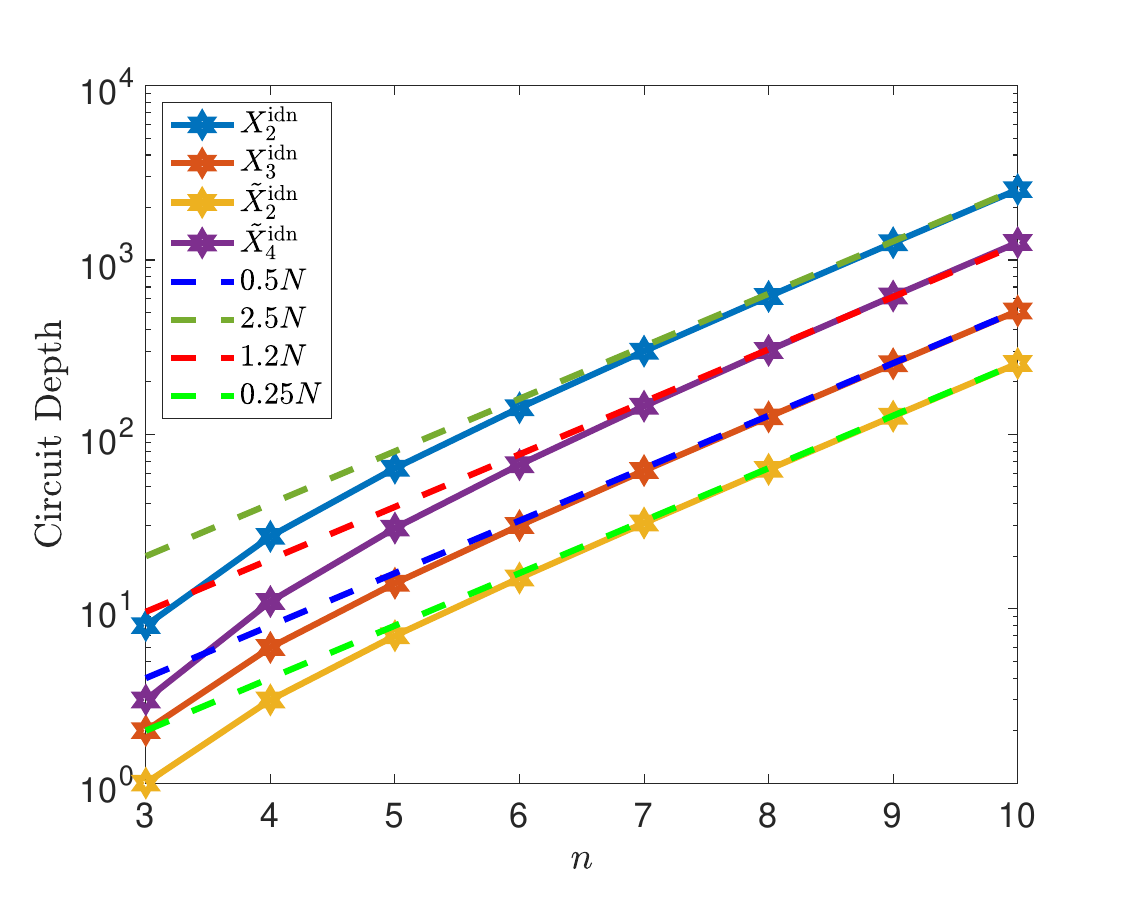}}
    \caption{Total circuit depth for concatenated unitaries corresponding to internal elements when all elements have the same lengths and diffusivity coefficients: (a) linear elements, (b) quadratic elements.}
    \label{fig:linear_concatenated_circuit_element_depths}
\end{figure}

Figure~\mbox{\ref{fig:linear_concatenated_circuit_element_depths}}a shows the total circuit depth (without transpilation) of the concatenated unitaries corresponding to the internal elements with an increasing number of qubits, when all linear elements have the same lengths and diffusivity coefficients. The circuit depth equals to $0.5 N$ for $X^{\mathrm{idn}}_{1}$ unitary and approaches $2.5 N$ for $X^{\mathrm{idn}}_{2}$ unitary.

It is important to note that the maximum number of unitaries that can be concatenated based on the idea of identical non-interacting elements is limited by the near-term gate fidelity of circuits with a depth of a few hundred in NISQ computers \mbox{\cite{zlokapa2023boundaries}}. Based on this limit on circuit depth, the number of identical non-interacting elements can be increased accordingly in the proposed approach, and this limit is rapidly increasing with advancements in quantum computing hardware.

\subsubsection{Quadratic elements}
The global stiffness matrix for the quadratic elements can be similarly represented with a reduced number of global element matrices using the concept of identical non-interacting elements. The boundary elements are again treated as identical non-interacting elements. The representation obtained for the global element stiffness matrix is shown below:
\begin{equation} 
\begin{aligned}
    \mathbf{K} & = \left( \frac{5 c^0}{2 h^0} + \displaystyle\sum_{e=1}^{n^{\mathrm{idn}}_{\mathrm{el}}-2} \frac{15 c^e}{3 h^e} + \frac{5 c^{n^{\mathrm{idn}}_{\mathrm{el}}-1}}{2 h^{n^{\mathrm{idn}}_{\mathrm{el}}-1}} \right) \: I + \frac{c^0}{h^0} \left( \frac{3}{2} \: Z_{0} - \frac{8}{3} \: X_{1} - \frac{4}{3} \: I_{1}^{-1} \right) \\
     & \quad \quad + \displaystyle\sum_{e=1}^{n^{\mathrm{idn}}_{\mathrm{el}}-2} \left( \frac{c^e}{h^e} \left( - \frac{8}{3} \: X^{\mathrm{idn}}_{2e} - \frac{8}{3} \: X^{\mathrm{idn}}_{2e+1} + \frac{1}{3} \: \Tilde{X}^{\mathrm{idn}}_{2e} \right) \right) \\ 
     & \quad \quad + \frac{c^{n^{\mathrm{idn}}_{\mathrm{el}}-1}}{h^{{n^{\mathrm{idn}}_{\mathrm{el}}-1}}} \left( \frac{3}{2} \: Z_{n_{\mathrm{el}}-1} - \frac{8}{3} \: X_{2(n_{\mathrm{el}}-1)} - \frac{4}{3} \: I_{2(n_{\mathrm{el}}-1)}^{-1} \right),
\end{aligned}
\end{equation}
where $X^{\mathrm{idn}}_{2e}$, $X^{\mathrm{idn}}_{2e+1}$, and $\Tilde{X}^{\mathrm{idn}}_{2e}$ are the unitaries for the $e^{\mathrm{th}}$ identical non-interacting quad element. The expressions for $X^{\mathrm{idn}}_{(\cdot)}$ and $\Tilde{X}^{\mathrm{idn}}_{2e}$ unitaries are:
\begin{eqnarray}
    X^{\mathrm{idn}}_{2e} &=& X_{2 e_0} X_{2 e_1} \cdots X_{2 e_{m}}, \\
    X^{\mathrm{idn}}_{2e + 1} &=& X_{2 e_0 + 1} X_{2 e_1 + 1} \cdots X_{2 e_{m} + 1}, \\
    \Tilde{X}^{\mathrm{idn}}_{2e} &=& \Tilde{X}_{2 e_0} \Tilde{X}_{2 e_1} \cdots \Tilde{X}_{2 e_{m}}
\end{eqnarray}
where $e_{(\cdot)}$ denotes the element number of non-adjacent quad elements and $X_{(\cdot)}$ denotes the corresponding circuit for that element. It must be noted that $e_{(\cdot)}$ can correspond to adjacent elements in $X^{\mathrm{idn}}_{2e}$ and $X^{\mathrm{idn}}_{2e+1}$ circuits, because of the structure of these unitaries. However, for consistency with $\Tilde{X}^{\mathrm{idn}}_{2e}$ circuits, we only consider non-adjacent quad elements in the identical non-interacting group element.

If all the elements have the same lengths and diffusion coefficients, the unitaries corresponding to the internal identical non-interacting element are obtained as:
\begin{eqnarray}
    X^{\mathrm{idn}}_{2} &=& X_{2} X_{4} \cdots X_{2 n_{\mathrm{el}} - 4}, \\
    X^{\mathrm{idn}}_{3} &=& X_{3} X_{5} \cdots X_{2 n_{\mathrm{el}} - 3}, \\
    \Tilde{X}^{\mathrm{idn}}_{2} &=& \Tilde{X}_{2} \Tilde{X}_{6} \cdots \Tilde{X}_{2 ( n_{\mathrm{el}}- 3)},  \\
    \Tilde{X}^{\mathrm{idn}}_{4} &=& \Tilde{X}_{4} \Tilde{X}_{8} \cdots \Tilde{X}_{2 ( n_{\mathrm{el}}- 2)}.
\end{eqnarray}
Note that for this simple example, the adjacent elements are also concatenated into an identical non-interacting element group for $X^{\mathrm{idn}}_2$ and $X^{\mathrm{idn}}_3$ circuits. If the boundary elements are combined similarly, the number of unitaries required to represent the global stiffness matrix is $8$ plus one additional unitary to make the number of unknowns $2^n$ for QC representation.

Figure~\mbox{\ref{fig:linear_concatenated_circuit_element_depths}}b shows the total circuit depth (without transpilation) of the concatenated unitaries corresponding to the internal elements with increasing number of qubits, when all quadratic elements have same lengths and diffusivity coefficients. The circuit depth approaches $0.5 N$ for $X^{\mathrm{idn}}_{3}$ unitary and $2.5 N$ for $X^{\mathrm{idn}}_{2}$ unitary, which is expected as these circuits are made from $X_{(\cdot)}$ circuits. While for other concatenated unitaries, the circuit depth approaches $1.2N$ for $\Tilde{X}^{\mathrm{idn}}_{2}$ unitary and $0.25N$ for $\Tilde{X}^{\mathrm{idn}}_{4}$ unitary.

To summarize the proposed method, the number of unitaries required to decompose the stiffness matrix for linear elements discretization is $n_{\mathrm{el}}^{\mathrm{idn}} + 1$, with $n_{\mathrm{el}}^{\mathrm{idn}}$ denoting the number of identical non-interacting element groups as defined in Sec.~\mbox{\ref{sec:reduction_unitaries}}. For quadratic elements, the number of unitaries required is $3 n_{\mathrm{el}}^{\mathrm{idn}} + 1$. We compare the number of unitaries required to decompose the stiffness matrix for the proposed method and the decomposition used in Ref.~\mbox{\cite{trahan2023variational}} in terms of matrix size $(N)$ for two extreme cases in Table~\mbox{\ref{tab:matrix_decomposition_comparison}}. The proposed approach has constant unitaries in the homogeneous case, albeit with an exponential increase in the circuit depth which is a limitation for NISQ devices, as discussed earlier.

\begin{table}[htb!]
\small
\centering
\begin{tabular}{|c|c|c|c|} 
 \hline
 Problem & Linear (Proposed) & Linear \cite{trahan2023variational} & Quadratic (Proposed) \\ \hline
 Completely Homogeneous  & 4 & N & 8  \\ 
 Completely Heterogeneous & N+2 & N  &  ${3}N/{2} + 1$  \\ 
 \hline
\end{tabular} 
\caption{Comparison of number of unitaries in matrix decomposition for two extreme problems.}
\label{tab:matrix_decomposition_comparison}
\end{table}

Multi-dimensional problems are expected to scale similarly to the one-dimensional problems. The number of unitaries in the proposed approach is upper bounded by $M^e \times n^{\mathrm{idn}}_{\mathrm{el}}$, where $M^e$ is upper bounded by the number of unique entries in the element stiffness matrix. For $d$ dimensional steady-state diffusion problem, the size of the element stiffness matrix is $2^d \times 2^d$ for linear elements and $3^d \times 3^d$ for quadratic elements. Since the element stiffness matrices are symmetric, $M^e$ is bounded by $2^{d}(2^d+1)/2$ for linear elements and $3^{d}(3^d+1)/2$ for quadratic elements. Notably, the ratio of degrees of freedom to the number of elements would be greater than $1$ for higher dimensional problems. However, it might not be large enough to counter the increase in $M^e$. Hence, the scaling of the number of unitaries for a completely heterogeneous problem would still be of $\mathcal{O}(N)$ for higher-dimensional problems.

\vspace{-0.3 cm}\subsection{Complexity of the algorithm} \label{sec:complexity}
\mbox{\citet{bravo2023variational}} performed a heuristic scaling of the VQLS algorithm for randomly generated matrices with random weights on combinations of Pauli matrices. Based on their numerical analysis, the number of iterations or the number of cost function evaluations scale linearly with the matrix condition number $(\kappa)$,  logarithmic scaling in $(1/\epsilon)$ (where $\epsilon = \frac{1}{2} \mathrm{Tr} \left| |u \rangle  \langle u| - | v\rangle \langle v| \right|$ is the trace distance between the exact $( | u \rangle)$ and numerical solution $( | v \rangle)$), and polylogarithmic scaling in problem size $N$. Combining the three scales, one can obtain the scaling as $\mathcal{O}\left(\kappa \log (1/ \epsilon) \left( \log N \right)^m\right)$, where $m \geq 1$ and is an integer. However, the scaling in $\epsilon$ assumes the components of the cost function for each unitary in the matrix decomposition can be measured by the corresponding Hadamard tests up to some precision $\epsilon$. In the case of finite sampling, the number of repetitions (or shots) in the VQLS algorithm would scale as $\mathcal{O}\left((\kappa/{\epsilon})^4\right)$ based on the scale of shot noise since the convergence criterion for the cost function is $\hat{C}_p < \epsilon^2 / \kappa^2$~\mbox{\cite{bravo2023variational}}. Given this desired tolerance for $\hat{C}_p$, the denominator $\langle \bs{\psi} | f \rangle$ and numerator $\langle \bs{\psi} | \bs{\psi} \rangle$ should be measured at least up to the desired precision. Each component of $\langle \bs{\psi} | f \rangle$ and  $\langle \bs{\psi} | \bs{\psi} \rangle$ 
 should be measured up to the same precision for the summation to be of that precision. Based on these heuristics, the overall complexity of the algorithm will approximately be  $ \mathcal{O}\left( L^2 P  \kappa \log (1/ \epsilon) \left( \log N \right)^m\right)$, where $P$ is the number of shots required to measure each of $ \mathcal{O}(L^2)$ circuits which would scale as $\mathcal{O}(\kappa^4 / \epsilon^4 )$. On the other hand, the classical complexity of the FEM for a sparse system, using the conjugate gradient as a linear solver, scales as $\mathcal{O} \left(N s \sqrt{\kappa} \log(1/\epsilon_{\mathrm{CG}})\right)$~\mbox{\cite{montanaro2016quantum}}, where $s$ is the number of non-zero entries in the rows and $\epsilon_{\mathrm{CG}}$ is the precision in `energy norm' $||\bs{v}||_{\mathrm{K}} := \sqrt{ \bs{v}^T\mathbf{K} \bs{v}}$. Therefore, if the number of circuits $(L^2)$ required for measuring the cost function scales as $O(1)$, then Q-FEM has an exponential advantage in $N$ compared to the classical FEM.

The decomposition of the stiffness matrix proposed in this work does not affect the complexity of the VQLS algorithm regarding the number of iterations for convergence. However, the scaling in the number of repetitions for all circuits $\mathcal{O}(L^2 P)$ would depend on the matrix decomposition strategy employed, as that determines the number of Hadamard tests needed for measuring the cost function. Moreover, the time required to execute each iteration of the cost function depends on the efficiency of the matrix decomposition, and its feasibility on the NISQ devices determined by the total circuit depth, single-qubit, and two-qubit gate counts. 

In what follows, we perform a heuristic scaling for the circuit characteristics of the two Hadamard tests shown in Fig.~\mbox{\ref{fig:VQLS_calcs}} for two bounding problems i.e., completely homogeneous and completely heterogeneous. In this analysis, we observe the scaling of the circuit depth, single-qubit, and two-qubit gate counts in the transpiled circuits of the two Hadamard tests, without accounting for the contribution due to the Hadamard gates (needed for measurement) and gates in the variational ansatz. Since, the circuits proposed for the unitaries in the stiffness matrix decomposition are multi-qubit gates, the circuits in the two Hadamard tests are transpiled into single and two-qubit gates. This is performed using IBM's \emph{Qiskit} library and the basis gates chosen for transpilation are the $H$, $X$, $Y$, $Z$, $P$, $Rx$, $Ry$, $Rz$, $CX$, and $CZ$ gates. Moreover, the optimization level $3$ is used for circuit transpilation in \emph{Qiskit}.

\begin{figure}[htb!]
    \centering
    \subfloat[][]{
    \includegraphics[scale=0.4]{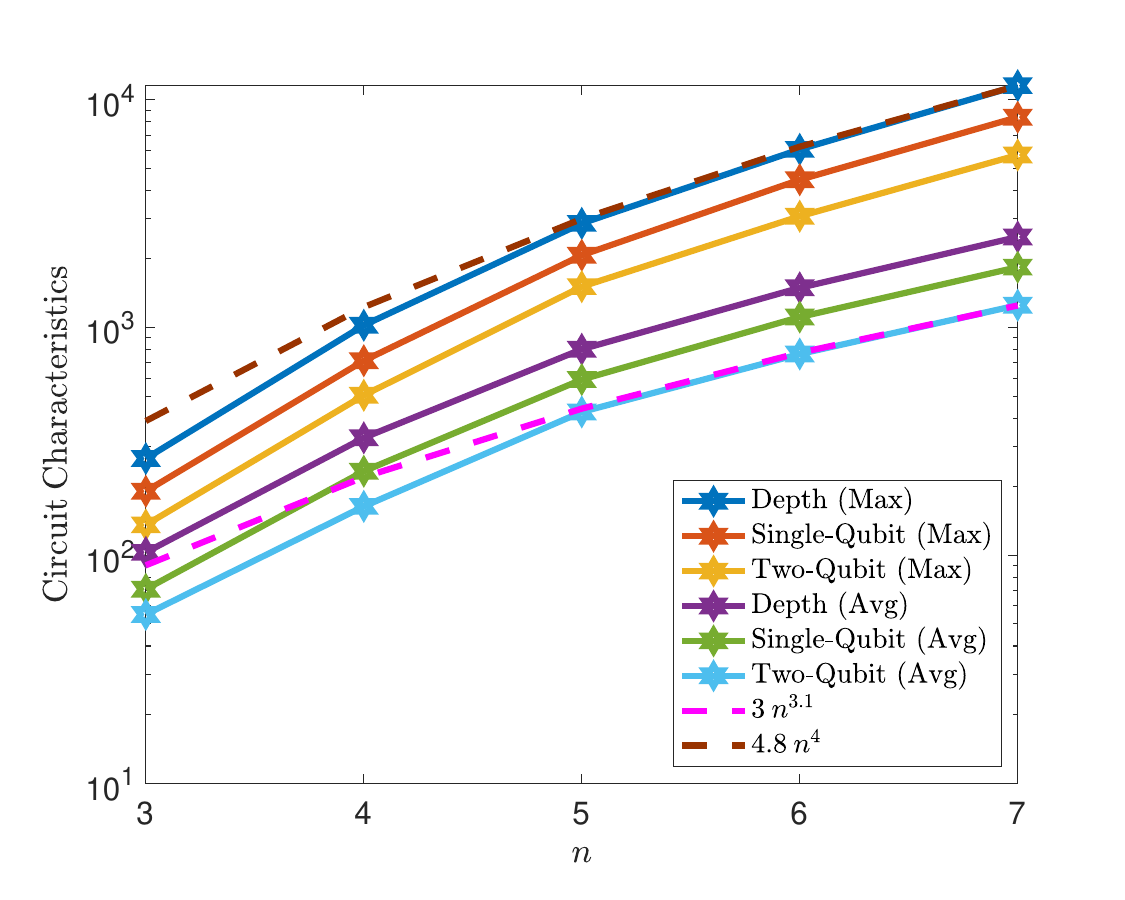}}
    \subfloat[][]{
    \includegraphics[scale=0.4]{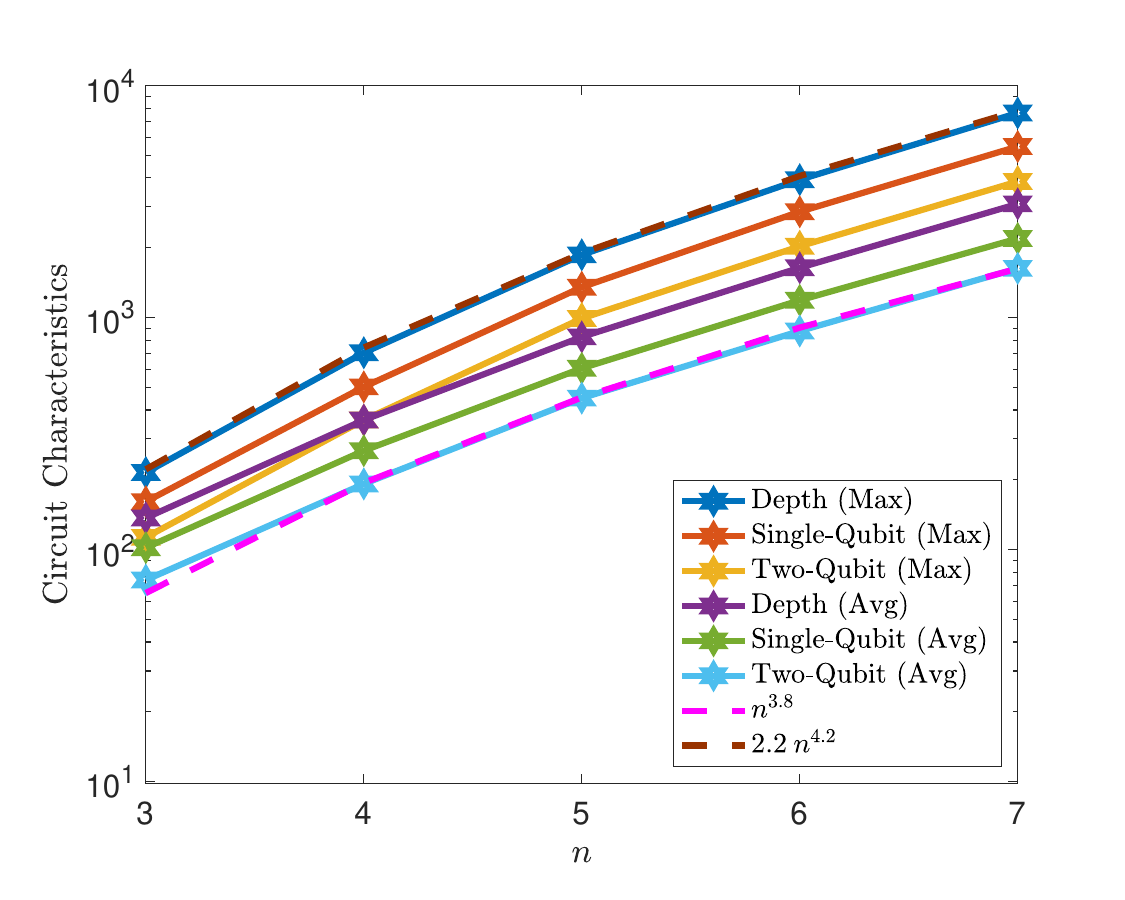}}
    \caption{The maximum and average number of circuit depth, single and two-qubit gate counts across all: (a) $\langle \bs{\psi} | \bs{\psi} \rangle$ circuits, and (b) $\langle \bs{\psi} | f \rangle$ circuits, for completely heterogeneous problem.}
    \label{fig:circuit_characteristics_heterogeneous}
\end{figure}

Figure~\mbox{\ref{fig:circuit_characteristics_heterogeneous}} reports the circuit characteristics for the completely heterogeneous problem and linear elements discretization $-$ where no unitaries can be concatenated. Since the circuit characteristics are different for various combinations of $n$ and $m$ in Eq.~\mbox{\eqref{costFunctionComponents}}b for $\langle \bs{\psi} | \bs{\psi} \rangle$ circuits and all possible $n$ values in Eq.~\mbox{\eqref{costFunctionComponents}}a for $\langle \bs{\psi} | f \rangle$ circuits, we report the maximum and average numbers of circuit depth, single and two-qubit gate counts across all circuits in Fig.~\mbox{\ref{fig:circuit_characteristics_heterogeneous}}. For $\langle \bs{\psi} | \bs{\psi} \rangle$ circuits, the scaling of various circuit characteristics asymptotes in between $\mathcal{O}(n^{3.1})$ and $\mathcal{O}(n^4)$, while for $\langle \bs{\psi} | f \rangle$ circuits, it scales between $\mathcal{O}(n^{3.8})$ and $\mathcal{O}(n^{4.2})$.

\begin{figure}[htb!]
    \centering
    \subfloat[][]{
    \includegraphics[scale=0.4]{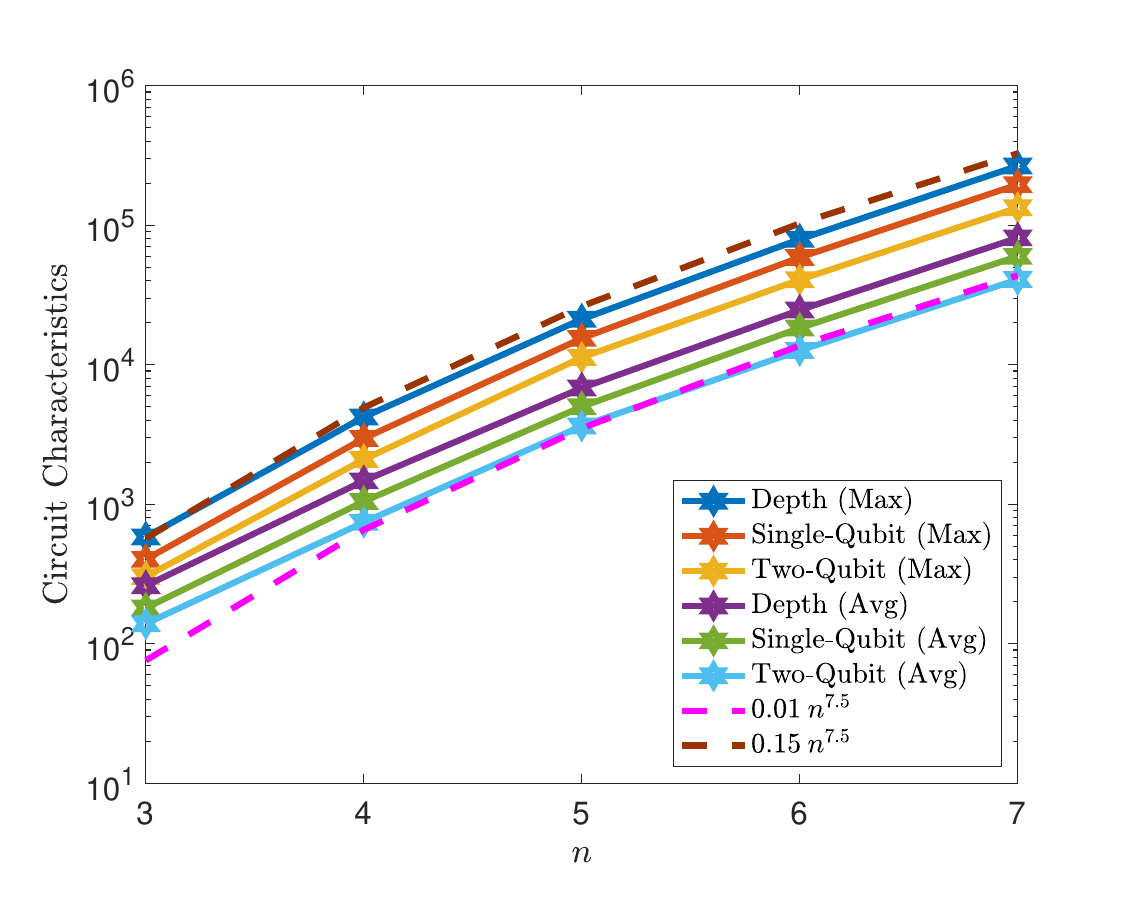}}
    \subfloat[][]{
    \includegraphics[scale=0.4]{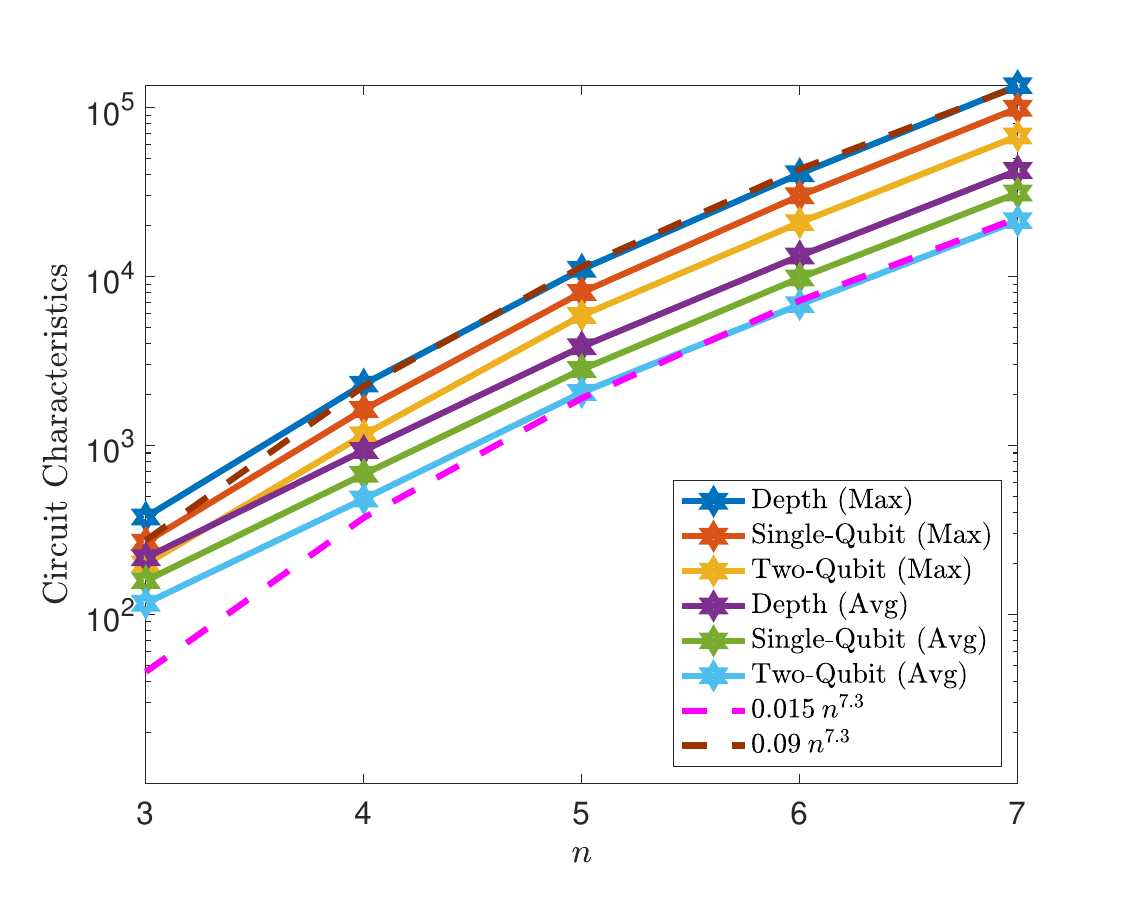}}
    \caption{The maximum and average number of circuit depth, single and two-qubit gate counts across all: (a) $\langle \bs{\psi} | \bs{\psi} \rangle$ circuits, and (b) $\langle \bs{\psi} | f \rangle$ circuits, for completely homogeneous problem.}
    \label{fig:circuit_characteristics_homogeneous}
\end{figure}

Similarly, Figure~\mbox{\ref{fig:circuit_characteristics_homogeneous}} reports the circuit characteristics for the completely homogeneous problem and linear elements discretization $-$ where unitaries for all identical non-interacting elements are concatenated. As evident in Fig.~\mbox{\ref{fig:circuit_characteristics_homogeneous}}, the scaling of various circuit characteristics asymptotes to $\mathcal{O}(n^{7.5})$ for $\langle \bs{\psi} | \bs{\psi} \rangle$ circuits, while for $\langle \bs{\psi} | f \rangle$ circuits, it scales as $\mathcal{O}(n^{7.3})$.

\vspace{-0.3cm} \subsection{Treatment for other boundary conditions} \label{sec:other_boundary_conditions}

The formulation can be extended to account for Neumann boundary conditions in a straightforward manner. The Neumann (natural) boundary condition contributes to the force vector alone. The force vector is assembled classically, including the boundary contributions, and represented using the quantum state vector representation. The construction of the stiffness matrix is performed as described above, but by discarding the boundary circuits used to impose homogeneous Dirichlet conditions. 

The non-homogeneous Dirichlet boundary conditions are implemented using the penalty method. Consider the strong form governing equation in Eq.~\eqref{strongFormEqn} with boundary conditions: $u(x_1) = \overline{u}, \: u(x_2) = 0$. For linear elements, the system of equations resulting from the penalty method implementation for the non-zero Dirichlet boundary conditions is shown below:
\begin{equation} \label{penaltyMatrix}
\begin{aligned}
\begin{bmatrix}
    \frac{c^0}{h^0} + P & -\frac{c^0}{h^0} &  0 &  &  &    \\ 
   -\frac{c^0}{h^0} & \frac{c^0}{h^0} + \frac{c^1}{h^1} & -\frac{c^1}{h^1} & &  &    \\
    0 & -\frac{c^1}{h^1} & \ddots & \ddots &  &   \\
       &  & \ddots &  \ddots &  \ddots & \\ 
       &  & &  \ddots &  \ddots &    \\
     &  &  &  &  &  -\frac{c^{n_{\mathrm{el}}-2}}{h^{n_{\mathrm{el}}-2}} \\ 
    &  &  &  & - \frac{c^{n_{\mathrm{el}}-2}}{h^{n_{\mathrm{el}}-2}} & \left( \frac{c^{n_{\mathrm{el}}-2}}{h^{n_{\mathrm{el}}-2}}  + \frac{c^{n_{\mathrm{el}}-1}}{h^{n_{\mathrm{el}}-1}}  \right)
\end{bmatrix} 
\begin{Bmatrix}
    u_0 \\ u_1  \\ \vdots \\ \vdots \\ \vdots \\  u_{N-2} \\ u_{N-1}
\end{Bmatrix} &= \begin{Bmatrix}
    f_0 + P \overline{u} \\ f_1  \\ \vdots \\ \vdots \\ \vdots \\  f_{N-2} \\ f_{N-1}
\end{Bmatrix},
\end{aligned}
\end{equation}
where $P$ denotes the penalty parameter. Naturally, the nodal degree of freedom associated with the non-homogeneous Dirichlet boundary condition is also considered an unknown, as implied in Eq.~\eqref{penaltyMatrix}. The global stiffness matrix is expressed in terms of unitary matrices as follows:
\begin{equation} 
    \mathbf{K} = \left( \frac{P}{2} + \sum_{e=0}^{n_{\mathrm{el}}-2} \frac{c^e}{h^e} + \frac{c^{n_{\mathrm{el}} - 1}}{2 h^{n_{\mathrm{el}} - 1}} \right) I - \frac{P}{ 2}  I_{0}^{-1}  -  \sum_{e=0}^{n_{\mathrm{el}}-2} \frac{c^e}{h^e} X_{e+1} - \frac{c^{n_{\mathrm{el}} - 1}}{2 h^{n_{\mathrm{el}} - 1}} I_{n_{\mathrm{el}}}^{-1} .
\end{equation}
The force vector is classically calculated with the addition of the contribution from the penalty term.  

\vspace{-0.3cm}\section{Numerical Verification} \label{sec:numerical_results}

The Q-FEM framework was implemented using IBM's \emph{Qiskit} software stack. Sequential least squares quadratic programming is used as the classical optimizer in the VQLS algorithm--leveraging the \emph{SLSQP} solver in the \emph{SciPy} optimization library. Several other optimization approaches exist, including other gradient-based optimizers as well as global optimizers (e.g. evolutionary algorithms). \emph{SLSQP} was chosen heuristically due to better-observed performance compared to other available techniques in the \emph{SciPy} optimization library. The cost function Jacobian is computed numerically using finite differences. 

Firstly, we consider the role of ansatz selection (from among those available in the literature) on problem convergence. Then, we assess the performance of the algorithm for linear and quadratic element discretizations as a function of increasing problem sizes. All the verification results are obtained using the \emph{Aer} simulator in the \emph{Qiskit} library, and without incorporating the noise model in the simulator. We also compare the performance of global and local cost functions for the linear elements example, and show the convergence challenges posed by the presence of barren plateaus and local minima as the problem size increases. 

\subsection{Ansatz testing}\label{sec:ansatz_testing}

The problem used in the ansatz testing is:
\begin{equation} \label{eq:ansatz_testing_ode}
    \frac{d^2 u (x)}{d x^2} + x = 0, \quad \forall x \in [0,1],
\end{equation}
subjected to homogeneous Dirichlet boundary conditions. We used 4 qubits in the main registry, which implies a discretization of the problem into 15 interior linear elements and 2 boundary linear elements. For the ansatz testing, we used two metrics to quantify ansatz characteristics: expressibility (Expr) and entanglement capability (ENT) \cite{meyer_global_2002, Sim_2019}. Expressibility aims to quantify how regularly the problem space is being sampled by the ansatz compared to a distribution of Haar random states. A high expressibility indicates the density of sampling of the problem space is relatively even throughout the sample space, while low expressibility indicates uneven sampling and lost information. Entanglement capability aims to describe the degree of quantum entanglement included in the ansatz.

We computed expressibility~\citep{Sim_2019} for the tested ansatzes by determining the Kullback-Liebler divergence between the probability distribution of fidelities of output states generated by randomly sampling the parameters of the ansatzes and the distribution of Haar random states calculated from an analytic expression~\cite{meyer_global_2002}:
\begin{eqnarray}
    \text{Expr} &=& D_{KL}(\hat{P}(F_{\bs{\theta},\bs{\phi}}) || P(F_{\text{Haar}})), \cr
    P(F_{\text{Haar}}) &=& (N-1)(1-F_{\text{Haar}})^{N-2}, \cr
    F_{\bs{\theta},\bs{\phi}} &=&  |\langle 0 | V^\dagger(\bs{\phi}) V(\bs{\theta})|0\rangle|^2, 
\end{eqnarray}
where fidelity $F_{\bs{\theta},\bs{\phi}}$ is the estimation of overlap between two pure states created by sampling the ansatz with two sets of random parameters $\bs{\theta}$, $\bs{\phi}$ \cite{Sommers_fidelity2005}. The fidelity $F_{\text{Haar}}$ is a uniform distribution of fidelities representing a purely random sampling of states. The probability density function $P(F_{\text{Haar}})$ has been derived in Ref.~\cite{M_Kus_1988}, and $N=2^n$ is the dimension of the Hilbert space. In expressibility computations below, the inner product of 4,000 pairs of ansatz-generated random states was used, which constituted the set of sampled states. Some variability in calculated fidelity was observed, so the values in Table \ref{tab:Ansatz_performance} are given as an average expressibility $\pm$ a standard deviation for 20 sets of sampled states per ansatz.

We estimated the entanglement capability for the tested ansatzes by averaging the Meyer-Wallach entanglement measure $Q$ \cite{Brennen2003,meyer_global_2002}:
\begin{eqnarray}
    \text{ENT} &=& \frac{1}{|S|} \sum_{\bs{\theta}_i\epsilon S} Q(V(\bs{\theta_i})|0\rangle), \cr 
    Q &=& 2 - \frac{2}{n} \sum_{k=1}^n \text{Tr}(\rho_k)^2,
\end{eqnarray}
where $\rho_k$ is the density matrix of the $k^\mathrm{th}$ qubit, $n$ is the number of qubits, and $S = \{\bs{\theta}_i \}$ is the set of sampled parameter vectors. ENT calculations employed 10,000 samples of circuit parameters.

\begin{figure}
    \centering
    \subfloat[][Ansatz 1: this manuscript.]{
    \includegraphics[scale = 0.45]{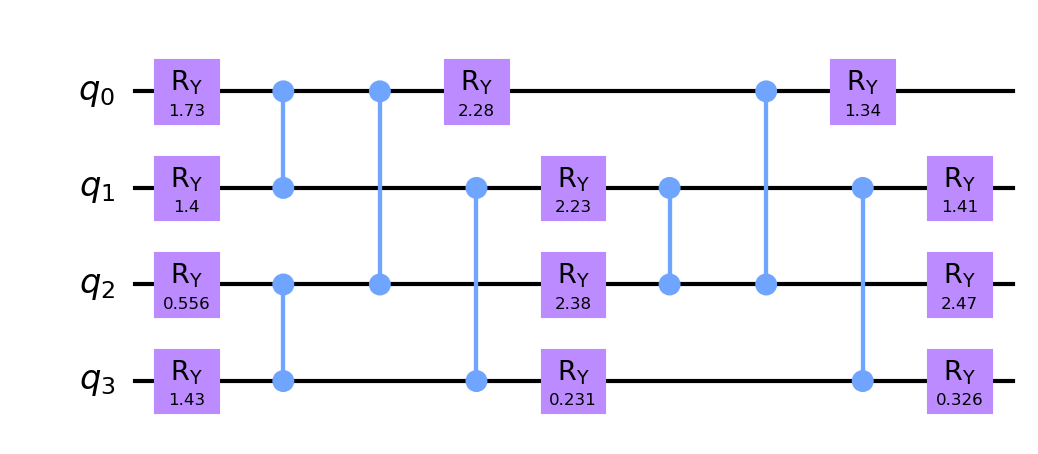}} 
    \subfloat[][Ansatz 2: \citet{bravo2023variational}.]{
    \includegraphics[scale = 0.45]{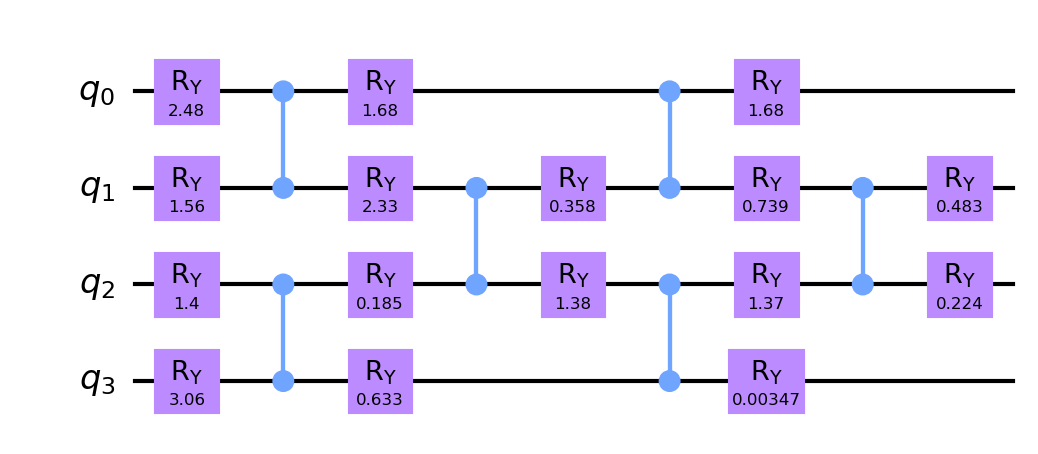}}
    \centering    \\
    \subfloat[][Ansatz 3: \citet{Sim_2019}.]{
    \includegraphics[scale = 0.45]{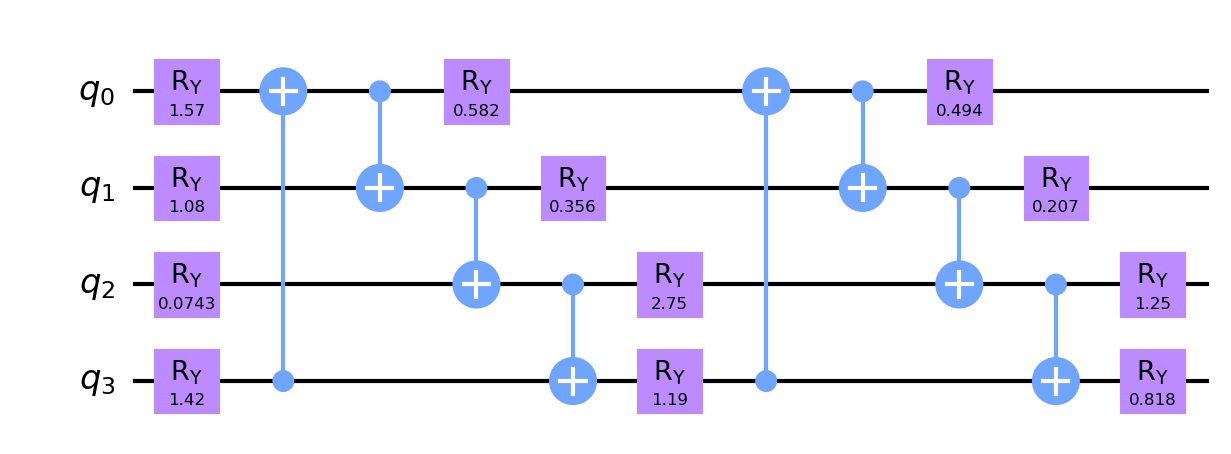}}
    \subfloat[][Ansatz 4: \citet{Sim_2019}.]{
    \includegraphics[scale = 0.45] {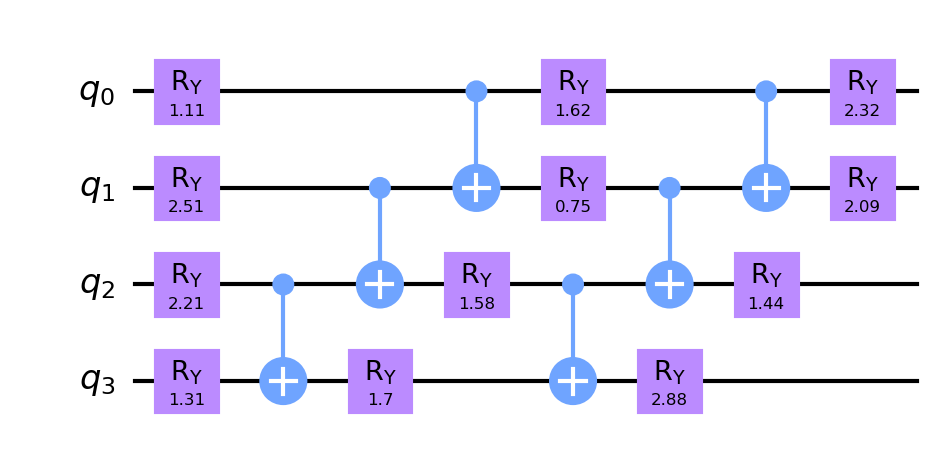}} \\ 
    \centering
    \caption{Four ansatzes tested in this study. Only the initial layer and one iterative layer are shown.}
    \label{fig:ansatzes}
\end{figure}

The structures of the tested ansatzes are shown in Fig.~\ref{fig:ansatzes}. The ansatzes differ from each other in the entanglement layers. In Q-FEM, the linear system of equations is real-valued, hence, ansatzes are chosen that exclusively use one-qubit $Ry$ rotation gates, which is commonly done in the literature such as in Ref.~\mbox{~\citep{bravo2023variational}} to limit the optimizer search to the solution subspace. It is important to note that limiting the search space to quantum states with only real amplitudes may impact the feasibility of achieving successful optimization. To achieve an entangled state, controlled Pauli gates ($CZ$ and $CX$) are employed in different patterns. While Ansatzes~2~\citep{bravo2023variational}, 3, and 4~\citep{Sim_2019} are taken from literature, Ansatz~1 is proposed herein to generalize the ansatz proposed in Ref.~\citep{bravo2023variational} to an odd number of qubits. We note that this ansatz features two-qubit gates that are not immediate neighbors of each other, which may pose difficulty in qubit arrangement in real quantum computers with imperfect connectivity.

Table~\ref{tab:Ansatz_performance} provides the overview of the ansatz performance testing and results. Each ansatz was tested in two, three, and four-layer configurations. Increasing the number of layers increases the number of ansatz parameters, hence the circuit cost~\cite{Sim_2019}. The ansatzes feature similar numbers of two-qubit gates per layer. For each configuration, Q-FEM was executed 20 times, starting from a random initial set of ansatz parameters to assess robustness in achieving a converged solution. The success rate (last column in Table~\ref{tab:Ansatz_performance}) refers to the number of simulations that successfully converged. Convergence is assessed when the cost function value at the minima found by the optimization algorithm is less than a tolerance value of $\hat{C}_p \leq 10^{-6}$. 

\begin{table}[htb!]
    \centering
    \small
    \begin{tabular}{|c|c|c|c|c|c|c|}
    \hline
    Ansatz & Ref & Layers & Expr & ENT & Parameters & Success (\%) \\
    \hline
    \hline
    \multirow{3}{*}{1} & \multirow{3}{*}{this manuscript} & 2 & .167$\pm$.008 & .765 & 12 & 0\\
    \cline{3-7}
     & & 3 & .134$\pm$.008 & .823 & 16 & 25\\
    \cline{3-7}
     & & 4 & .126$\pm$.010 & .847 & 20 & 100\\
    \hline
    \hline
    \multirow{3}{*}{2} & \multirow{3}{*}{\citet{bravo2023variational}} & 2  & .193$\pm$.014 & .502 & 16 & 0\\
    \cline{3-7}
    &  & 3 & .151$\pm$.012 & .684 & 22 & 100\\
    \cline{3-7}
    &  & 4 & .138$\pm$.009 & .684 & 28 & 100\\
    \hline
    \hline
    \multirow{3}{*}{3} & \multirow{3}{*}{\citet{Sim_2019}} &2 & .142$\pm$.009 & .881 & 12 & 0\\
    \cline{3-7}
     & &3 & .125$\pm$.007 & .900 & 16 & 100\\
     \cline{3-7}
     & &4 & .124$\pm$.008 & .896 & 20 & 100\\
    \hline
    \hline
    \multirow{3}{*}{4} & \multirow{3}{*}{\citet{Sim_2019}} & 2 & .170$\pm$.011 & .779 & 12 & 5\\
    \cline{3-7}
     & & 3 & .136$\pm$.009 & .838 & 16 & 30\\
     \cline{3-7}
    & & 4 & .130$\pm$.011 & .856 & 20 & 95\\
    \hline
    \end{tabular}
    \caption{Results of the ansatz testing.}
    \label{tab:Ansatz_performance}
\end{table}

A lower Expr value indicates a higher level of expressibility, whereas a higher value of ENT (unity as maximum) indicates high entanglement capability. Expressibility and entanglement generally show a consistent and improving trend with increasing layer depths. The success rate also increases with layer depth for all ansatzes tested. Among the four ansatzes tested, Expr and ENT values are most favorable for Ansatzes 1 and 3, respectively. Ansatz 3 demonstrates the highest values of success rate with a low number of parameters. Higher layer depth can lead to very flexible circuits capable of producing low entanglement states, which may begin to lower ENT~\cite{Sim_2019}, as appears to be the case in Ansatz 3. Expr saturation is also observed in Ansatz 3 behavior in the transition from 3 to 4 layer depth, a trend noted for a similar ansatz in Ref.~\cite{Sim_2019}. When the number of parameters is less than or equal to 16, the success rate is generally low. This trend is attributed to the observation that the size of the system of linear equations is 16, hence a 16-parameter system (unless degenerate) may have a solution, and ansatzes with larger than 16 parameters sample the space more effectively and increase the chances of finding the correct solution. We note that the presence of quantum entanglements results in a highly nonlinear cost function. It is, therefore, possible to achieve an optimal solution with less than 16 parameters but with a low success rate. Note that Ansatz $4$ achieved a $5$\% success rate with only $12$ parameters. Prior studies demonstrated well-behaved convergent examples in the context of VQLS, where the linear system size is of orders of magnitude larger than the number of ansatz parameters~\citep{bravo2023variational}. However, the condition numbers of the Ising-inspired systems tested in the prior studies are near unity, whereas those that correspond to problems investigated in this manuscript are substantially higher, as further discussed in Section~\ref{sec:problem_scaling}. 

\subsubsection{Convergence characteristics}

\begin{figure}[htb!]
    \centering
    \includegraphics[scale=.65]{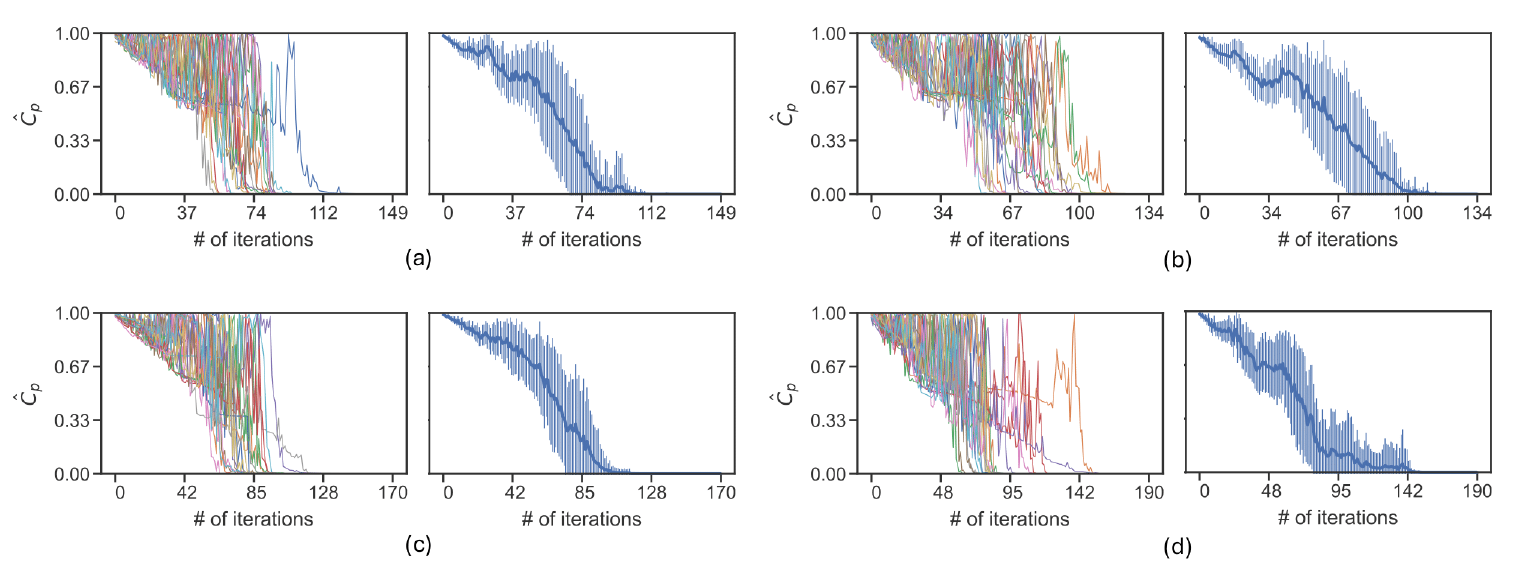}
    \caption{Evolution of the cost function with iterations using (a)~Ansatz 1, (b)~Ansatz 2, (c)~Ansatz 3, and (d)~Ansatz 4. The left figure in each sub-figure shows various runs and the right figure shows the evolution of mean and standard deviation across the ensemble of runs.}
    \label{fig:ansatzes_results}
\end{figure}

\begin{figure}[htb!]
    \centering
    \subfloat[][($\theta_1, \theta_2$)]{
    \includegraphics[scale=0.5]{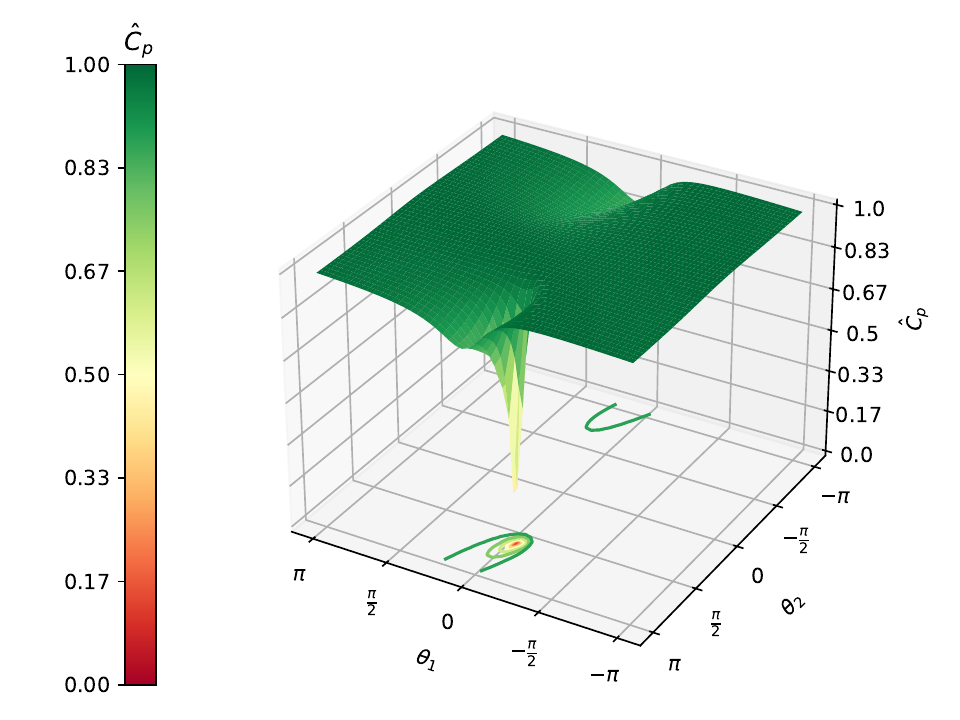}}
    \subfloat[][($\theta_8, \theta_9$)]{
    \includegraphics[scale=0.5]{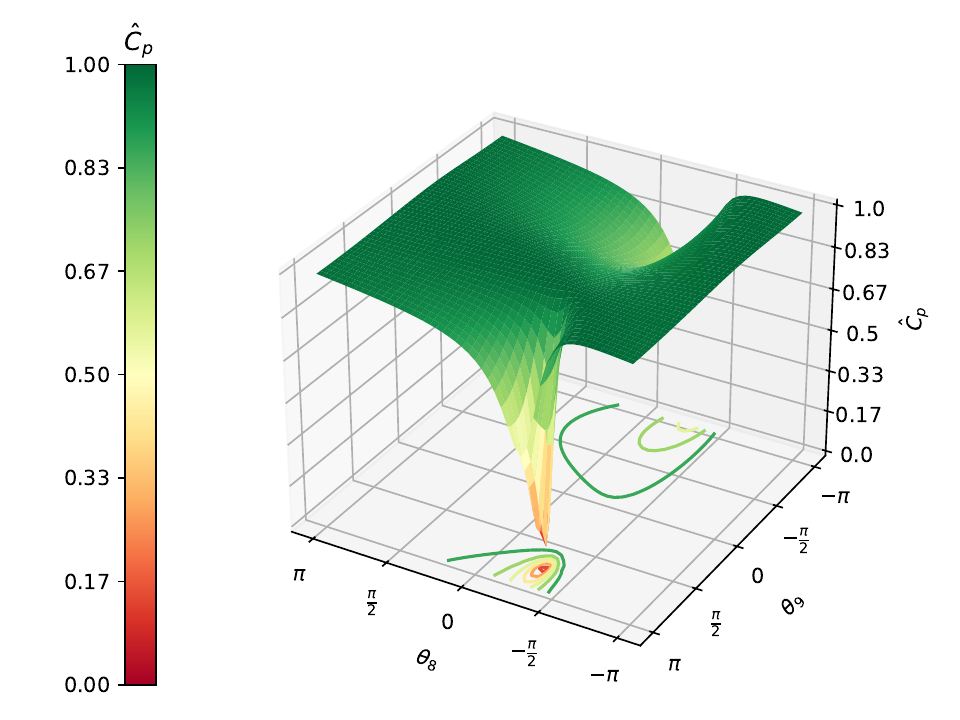}}
    \caption{The field plot for cost function space obtained by varying only two parameters and all other parameters correspond to the minima solution for a 3-qubit linear element discretization example.}
    \label{fig:cost_function_space_plot}
\end{figure}

Figure~\ref{fig:ansatzes_results} shows the evolution of the cost function as a function of optimization iterations for each of the ansatzes tested in this study. The plots show 20 runs of tested Ansatzes $1-4$ with a layer depth of $3$ for Ansatzes $2$, $3$ and a layer depth of $4$ for Ansatzes $1$, $4$. A highly oscillatory pattern towards convergence is evident regardless of the initial state and the choice of the ansatz. This pattern was previously observed in other works as well~\citep{trahan2023variational}. We attribute this highly oscillatory behavior to the cost function landscape. Figure~\ref{fig:cost_function_space_plot} illustrates the cost function topology, which is obtained by parametric analysis of the cost function near the global minimum. The topology was calculated for a 3-qubit system by varying two ansatz parameters while keeping all other parameters fixed at the global minima solution. The cost function topology is marked by barren plateaus with multiple local minima solutions located in sharp valleys (deep gorges). During the optimization iterations, low-cost function gradients computed at barren plateaus resulted in large optimization steps and increased cost function values. Fortunately, large step sizes allow for globally probing the cost function landscape and eventually reaching the minimum. In contrast, step size control (e.g., using line search) results in probing the local vicinity of the initial guess and does not necessarily improve the success rate despite achieving more monotonic behavior. The differential evolution algorithm was shown to be resilient to vanishing gradients and local minima in the 1-d Ising chain problem for variational quantum algorithms~\mbox{\cite{failde2023using}}. Using the differential evolution algorithm for the optimization problems in QFEM may be beneficial to avoid local minima and barren plateaus, but we leave such an exercise for future work.

Ansatzes $1$ and $3$ converge successfully in approximately $100$ iterations and exhibit a similar pattern, while Ansatz $2$ converged in about $120$ iterations. Ansatz $4$ required substantially more iterations in certain cases, where the optimization iterations reached plateau regions before achieving convergence. The statistical behavior (i.e., the evolution of mean and the standard deviation) shown in Fig.~\ref{fig:ansatzes_results} demonstrates a relatively monotonic convergence behavior in the aggregate.  

\subsection{Effect of boundary and force conditions} \label{section:linear_element_tests}
We proceed with the assessment of the capabilities of Q-FEM by considering linear finite element discretization of the problem in a slightly generalized form $u''(x) + b(x) = 0$ using a variety of boundary conditions. The grid and the coefficients are taken to be uniform in the examples shown in this section. Convergence criterion used for the global cost function is $\hat{C}_p \leq 0.5\times10^{-5}$.

\begin{figure}[htb!]
\centering
    \includegraphics[scale=.5]{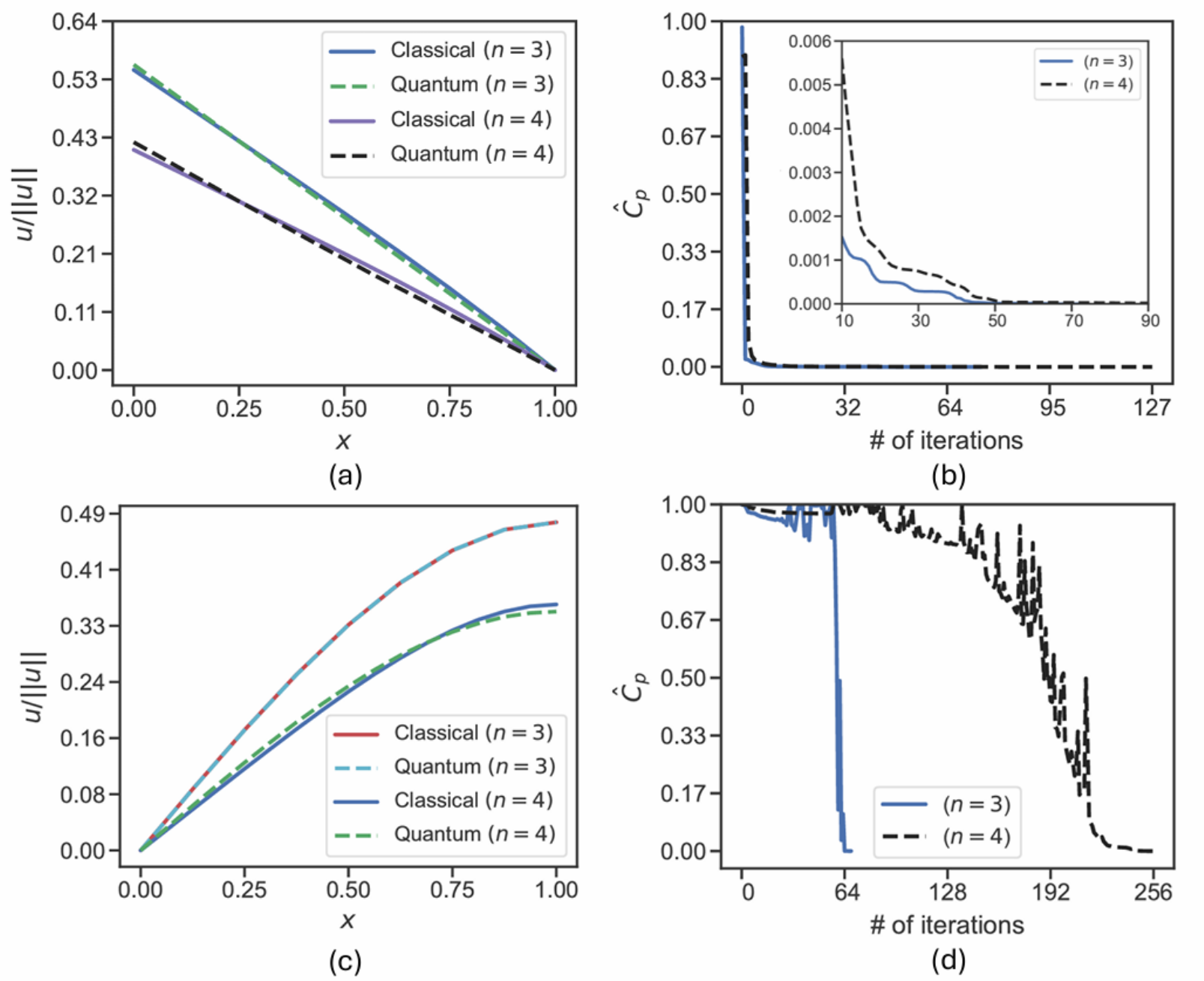}
    \caption{(a) Comparison of Q-FEM and the classical FEM solutions with boundary conditions $u(0)=1$, $u(1)=0$, and $b(x)=x^2$. The 3 and 4 qubit solutions are obtained using Ansatz 1. (b)~Corresponding evolution of the cost function. Inset figure provides zoom-in view. (c) Comparison of Q-FEM and the classical FEM solutions with boundary conditions $u(0)=0$, $u'(1)=0$, and $b(x)=x$. The 3 and 4 qubit solutions are obtained using Ansatz 4. (d)~Corresponding evolution of the cost function.}
    \label{lowQubitLinearExamples}
\end{figure}

Figure~\ref{lowQubitLinearExamples}a-b shows the results when the boundary conditions are set to $u(0) = 1$ and $u(1) = 0$, and the forcing function is taken as $b(x)=x^2$. The penalty method is used to implement the non-zero Dirichlet boundary condition as discussed in Section~\ref{sec:other_boundary_conditions} and for this example, the penalty parameter used is $P=100$. The comparison of the normalized solution field obtained by Q-FEM and the classical FEM for $n=3$ and $n=4$ cases are shown in Fig.~\ref{lowQubitLinearExamples}a. Q-FEM results are obtained by setting ansatz layer depth to 2 and 4 for $n=3$ and $n=4$, respectively. We note that the 3-qubit and 4-qubit results do not overlap due to the normalization with respect to the Euclidian norm of the solution vector. Each Q-FEM simulation was repeated 25 times with random initial parameters using Ansatz 1, and 100\% of these repeat runs resulted in convergence for 3-qubit. In contrast, the convergence probability was 92\% for the 4-qubit example. Figure~\ref{lowQubitLinearExamples}b shows the typical convergence behavior of Q-FEM with Ansatz 1, where the 3-qubit and 4-qubit simulations converged with 75 and 128 iterations, respectively. The convergence behavior of this example is substantially more regular, exhibiting fewer oscillations and rapid initial convergence compared with Fig.~\ref{fig:ansatzes_results}, while the average number of iterations to convergence is similar. This observation points to the influence of the boundary condition on the topology of the cost function. We further note that the condition number of the stiffness matrix for the current example (that uses the penalty method to impose non-homogeneous Dirichlet boundary condition) is higher than that for the case with homogeneous Dirichlet boundary conditions reported in Fig.~\ref{fig:ansatzes_results}. 


Figure~\ref{lowQubitLinearExamples}c shows the results when the boundary conditions are set to $u(0) = 0$ and $u'(1) = 0$. The forcing function is linear $b(x) = x$ in this example. Q-FEM results are obtained by setting ansatz layer depth to 4 for both $n=3$ and $n=4$ cases (layer depth of 2 did not result in a high probability of convergence in $n=3$ case). Figure~\ref{lowQubitLinearExamples}d shows the typical convergence behavior of Q-FEM with Ansatz 4, where the 3-qubit and 4-qubit simulations converged with 68 and 256 iterations, respectively. 




\subsection{Problem scaling} \label{sec:problem_scaling}

%

The performance of Q-FEM with an increasing number of qubits is assessed using the problem defined in Eq.~\eqref{eq:ansatz_testing_ode}. Ansatz 1 was used in the simulations. Convergence is assessed when the cost function value is less than a tolerance value of $\hat{C}_p \leq 2.5 \times 10^{-3}$. As the problem size increases, the optimization problem becomes increasingly higher-dimensional as the size of the parameter vector also increases. Numerical testing revealed that the cost function exhibits barren plateaus, leading to the optimization problem either converging to a local minimum or failing to converge within a reasonable number of iterations, even with global optimizers. To solve problems with more than $4$ qubits, we use the classical FEM solution to obtain a good initial guess for the parameters in the ansatz. Previous work, such as \cite{renaud2024quantum}, has reported the generation of good initial guesses for various problems using quantum computing. This underscores the importance of finding a good initial guess, highlighting the need to design better ansatzes that span the solution space more effectively with as few parameters as possible.

\begin{figure}[htb!]
    \centering
    \includegraphics[width=8cm]{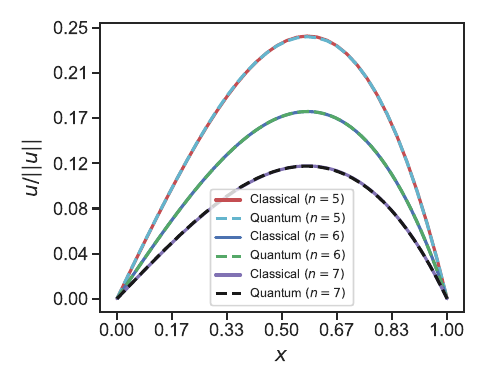}
    \caption{The comparison of Q-FEM solution with the classical FEM solution for homogeneous coefficients and same element lengths problem with increasing qubits.}
    \label{fig:linear_qfem_solution_increasing_qubits_comparison}
\end{figure}

\begin{table}[htb!]
    \small
    \centering
    \begin{tabular}{|c|c|c|}
    \hline 
     \# of qubits & Ansatz layer depth & \# of parameters \\ \hline
    3 & 2  & 9 \\
    4 & 4 & 20 \\
    5 & 6 & 42 \\  
    6 & 13 & 84 \\
    7 & 22 & 161 \\
    \hline
    \end{tabular}
    \caption{Layer depth and parameters in the ansatz to obtain a correct solution with Q-FEM framework for increasing qubits.}
    \label{tab:ansatz_layer_depth_with_no_of_qubits_linear}
\end{table}

\begin{figure}[htb!]
    \centering
    \includegraphics[scale = 0.5]{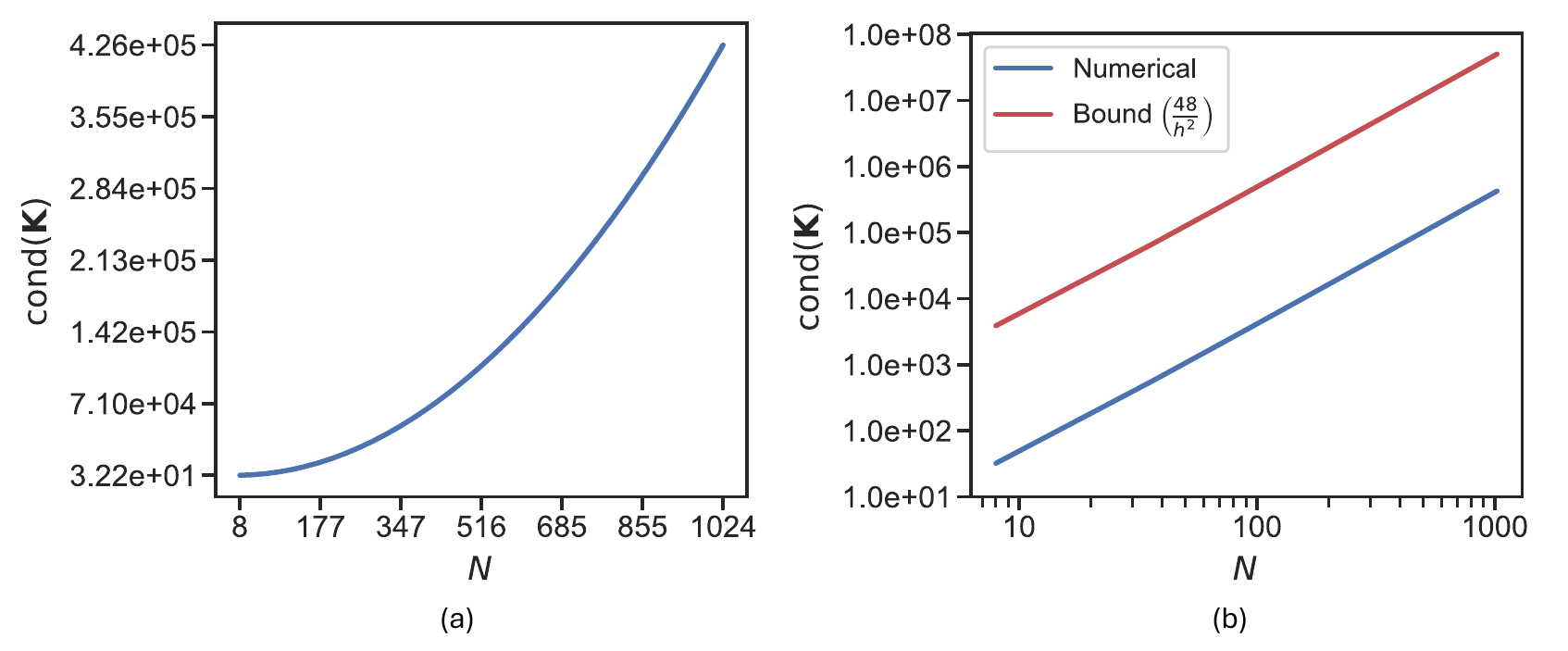}
    \caption{(a) The condition number of the matrix with increasing system size. (b) Comparison of numerically obtained condition number of the matrix and its theoretical bound with system size.}
    \label{fig:condition_number_linear_element}
\end{figure}

Using the generated initial guesses for parameters in the ansatz, we obtain solutions with the Q-FEM framework for increasing problem sizes, which are in good agreement with the classical solution, as shown in Fig.~\ref{fig:linear_qfem_solution_increasing_qubits_comparison}. Achieving the correct solution for increasing problem sizes requires the number of parameters in the variational ansatz to scale with the number of unknowns, as shown in Table \ref{tab:ansatz_layer_depth_with_no_of_qubits_linear}. In the classical solution of a full-rank linear system, the unknown variables are equal to the number of equations. For VQLS, the number of unknowns can be less, equal, or more than the number of equations, depending on the specific structure of the nonlinear variational ansatz and its ability to express the correct solution of the specific linear system problem. For the current example, using a larger number of parameters than the number of equations increases the probability of convergence and is required in practice to obtain the correct solution as indicated in Table~\ref{tab:ansatz_layer_depth_with_no_of_qubits_linear}. However, for the Ising-inspired example problems discussed in Ref.~\cite{bravo2023variational}, the number of parameters in the ansatz scales linearly with the number of qubits, which is equivalently logarithmic scaling in the number of equations. We speculate that the reason for this scaling is the specific structure of the linear system, which allows the variational ansatz to span the solution space more effectively with fewer parameters. Moreover, the condition number of the global stiffness matrix ($\mathrm{cond}(\mathbf{K})$) increases with the problem size, as shown in Fig.~\ref{fig:condition_number_linear_element}a. The numerically obtained $\mathrm{cond}(\mathbf{K})$ for Eq.~\eqref{eq:ansatz_testing_ode} scales as $\sim N^2$, which matches with the theoretical scaling for the upper bound of the condition number, $\mathrm{cond}(\mathbf{K}) \leq 48/h^2 = 48 (N+1)^2$ where $h = 1/(N+1)$ is the element length, deduced in \cite[Sec.~5.2]{strang_fem_analysis}. This is evident from the slope of the log-log plot between $\mathrm{cond}(\mathbf{K})$ (numerically obtained and the theoretical bound)  and $N$ shown in Fig.~\ref{fig:condition_number_linear_element}b. However, this is not observed in the examples considered in \cite{bravo2023variational}, where the condition number remains constant regardless of the matrix size. The increasing condition number with system size escalates the time complexity of the optimization problem and raises the potential for encountering barren plateaus and local minima solutions. Some mitigation strategies to tackle the convergence issues could be: (1) solving the preconditioned linear system~\mbox{\cite{clader2013preconditioned}}, (2) ansatz design informed by the specific linear system, (3) using differential evolution  for optimization in the VQLS algorithm~\mbox{\cite{failde2023using}}, and (4) posing the second-order differential equation as a system of first-order equations and using least-squares finite element method to reduce the stiffness matrix condition number~\mbox{\cite{zhou2021improving}}.

Table~\mbox{\ref{tab:circuit_characteristics_problem_scaling}} reports the maximum number of circuit depth, one-qubit, and two-qubit gate counts for the homogeneous problem defined in Eq.~\mbox{\eqref{eq:ansatz_testing_ode}} with same element lengths, among all possible $ \langle \bs{\psi}| \bs{\psi} \rangle$ and $\langle \bs{\psi}| f \rangle$ circuits. Both the circuits are transpiled including the Hadamard gates (for measurement) and variational ansatz gates with appropriate circuit depth as given in Table~\mbox{\ref{tab:ansatz_layer_depth_with_no_of_qubits_linear}}, and the basis gates and optimization level chosen for tranpilation is the same as reported earlier in Sec.~\mbox{\ref{sec:complexity}}. It must be noted that the total number of unitaries for the stiffness matrix decomposition is $4$ for all qubits, as all possible unitaries are concatenated.

\begin{table}[htb!]
    \small
    \centering
    \begin{tabular}{|c|c|c|c|}
    \hline 
     \# of qubits & Circuit depth & One-qubit gates & Two-qubit gates\\ \hline
    3 & 579 & 403 & 306 \\ 
    4 & 4223 & 2976 & 2102 \\ 
    5 & 21182 & 15466 & 11286 \\  
    6 & 79274 & 58749 & 40685 \\
    7 & 265594 & 195388 & 133496 \\ 
    \hline
    3 & 434 & 320 & 234 \\ 
    4 & 2471 & 1751 & 1277 \\ 
    5 & 11354 & 8270 & 6117 \\
    6 & 41428 & 30613 & 21472 \\ 
    7 & 136478 & 100190 & 69043 \\  
    \hline
    \end{tabular}
    \caption{The maximum numbers of total circuit depth, one-qubit, and two-qubit gate count among all circuits (transpiled) for homogeneous coefficients and same element lengths problem with increasing qubits. The second to sixth rows correspond to $ \langle \bs{\psi}| \bs{\psi} \rangle$ circuits, while the last five rows correspond to $ \langle \bs{\psi}| f \rangle$ circuits.}
    \label{tab:circuit_characteristics_problem_scaling}
\end{table}


\subsection{Local cost function}\label{section:local_cost_function}
Previous studies, such as Ref.~\mbox{\cite{bravo2023variational}}, have shown that the barren plateau problem and associated optimization challenges can be partially mitigated by using a local cost function instead of a global cost function based on the projection operator. In this section, we test the local cost function within the Q-FEM framework for the problem defined in Eq.~\mbox{\eqref{eq:ansatz_testing_ode}}, subject to homogeneous Dirichlet boundary conditions. Ansatz $1$ is used for this example, with the layer depth of $2$ and $4$ for $n=3$ and $n=4$ qubits, respectively. The performance of the local cost function is compared to that of the global cost function, both initialized with the same starting parameters and assessed using a convergence criterion of $10^{-5}$.

\begin{figure}[htb!]
    \centering
    \includegraphics[width=8cm]{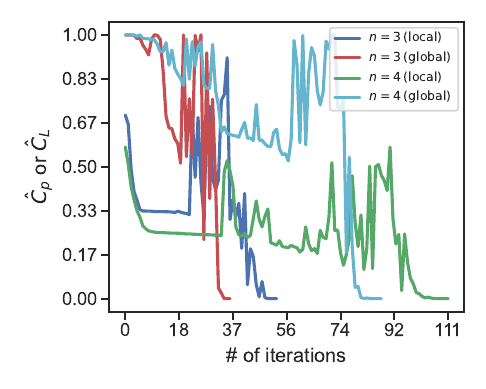}
    \caption{The comparison of the evolution of local and global cost functions for a homogeneous problem with boundary conditions $u(0) = u(1) = 0$, and $b(x) = x$. }
    \label{fig:local_cost_convergence}
\end{figure}

As shown in Fig.~\mbox{\ref{fig:local_cost_convergence}}, for both $n=3$ and $n=4$ qubits, the global cost function converges in fewer iterations than the local cost function. The lack of improvement with the local cost function can be attributed to the required ansatz layer depth needed to resolve the solution space, which scales as $\mathcal{O}(n^a)$ (with $ 0 < a < 2$), as shown in Section \mbox{\ref{sec:problem_scaling}}. Additionally, since the local cost function requires evaluating a more significant number of circuits, which are deeper as discussed in Ref.~\mbox{\cite{bravo2023variational}}, without yielding much convergence benefits in the tested cases, the global cost function appears to be more effective for solving the linear systems arising from FEM discretizations performed in this study.

\subsection{Quadratic elements} \label{section:quadratic_element_tests}

In this section, we consider a heterogeneous domain such that $b(x)=x$, $x_1=0$, $x_2=1$ and $c(x) = 1.5$ for $ x \in [0, 0.41)$ and $c(x) = 2.0$ for $x \in (0.41, 1]$ in Eq.~\ref{strongFormEqn}. We discretize the domain into $n_{el}= 4$ and $n_{el}= 8$ quadratic elements, and the corresponding number of qubits are $n = 3$ and $n=4$, respectively. The number of internal nodes for $n=3$ is $N = 7$, and for $n=4$ it is $N=15$. As stated previously, we add an auxiliary degree of freedom to represent the stiffness matrix and the force vector in the quantum computer; and without loss of any generality, the corresponding auxiliary entry in the force vector is taken to be $0$. Tables~\ref{tab:quadratic_3_qubits_discretization} and \ref{tab:quadratic_4_qubits_discretization} show the element lengths and coefficients taken for each element in FEM discretization for $n=3$ and $n=4$, respectively. Ansatz $1$ is used in this example, and convergence criterion used is $\hat{C}_p \leq 10^{-3}$. 

\begin{table}[htb!]
    \small
    \centering
    \begin{tabular}{|c|c|c|c|c|}
    \hline 
      & $e=1$ & $e=2$ & $e=3$ & $e=4$ \\ \hline
    $h^e$ &  0.21 & 0.2 & 0.235 & 0.355 \\  
    $c^e$ &  1.5 & 1.5 & 2.0 & 2.0  \\
    \hline
    \end{tabular}
    \caption{Element lengths and coefficients for a 3-qubit quadratic element discretization.}
    \label{tab:quadratic_3_qubits_discretization}
\end{table}

\begin{table}[htb!]
    \small
    \centering
    \begin{tabular}{|c|c|c|c|c|c|c|c|c|}
    \hline 
      & $e=1$ & $e=2$ & $e=3$ & $e=4$ & $e=5$ & $e=6$ & $e=7$ & $e=8$ \\ \hline
    $h^e$ &  0.105 & 0.105 & 0.1 & 0.1 & 0.125 & 0.125 & 0.17 & 0.17 \\  
    $c^e$ &  1.5 & 1.5 & 1.5 & 1.5 & 2.0 & 2.0 & 2.0 & 2.0 \\
    \hline
    \end{tabular}
    \caption{Element lengths and coefficients for a 4-qubit quadratic element discretization.}
    \label{tab:quadratic_4_qubits_discretization}
\end{table}

\begin{figure}[htb!]
    \centering
    \includegraphics[scale = 0.5]{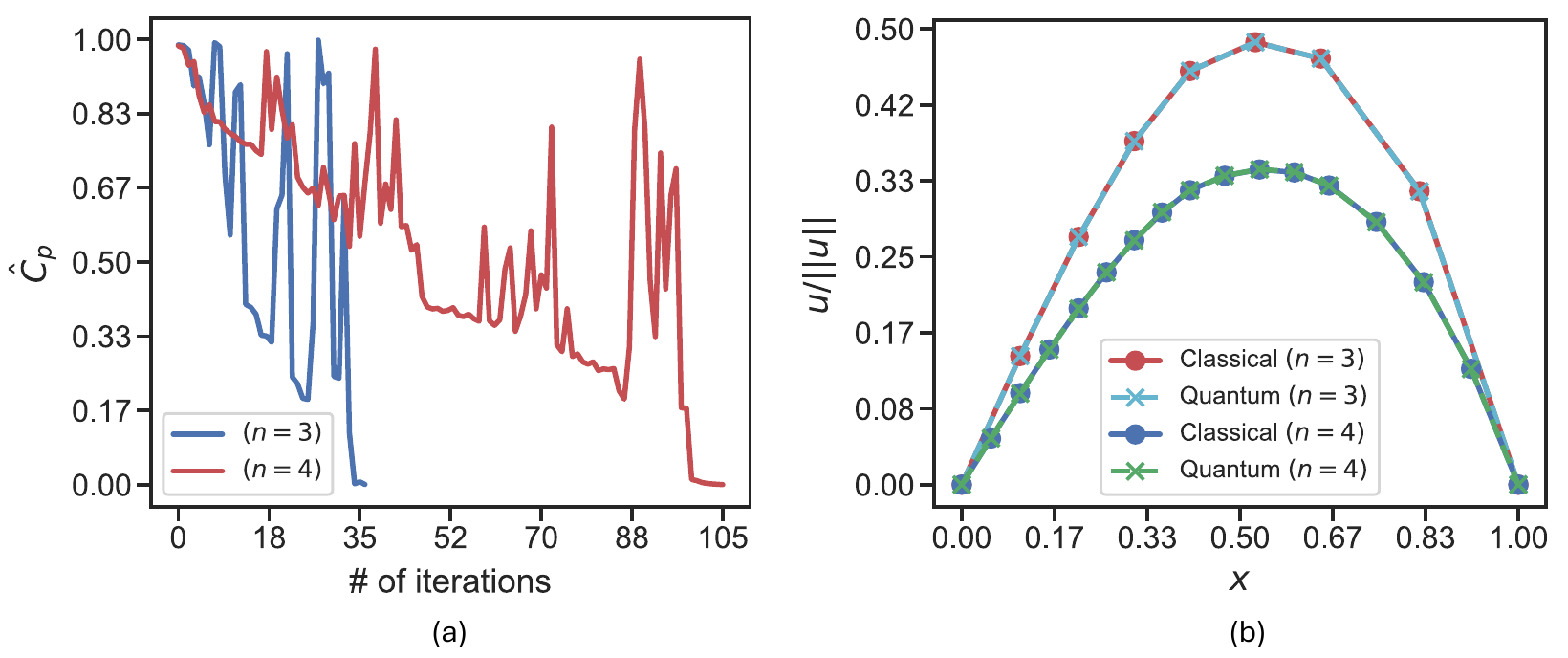}
    \caption{(a) The cost function vs the number of iterations for different qubits with quadratic element discretization. (b) The comparison of Q-FEM solution with the classical FEM solution.}
    \label{fig:cost_function_results_quadratic_elements}
\end{figure}

For $n=3$ and $n=4$ examples, the ansatz layer depth is set to 2 and 4, respectively. Figure \ref{fig:cost_function_results_quadratic_elements}a shows the value of the cost function with the number of iterations for both examples. As evident in Fig.~\ref{fig:cost_function_results_quadratic_elements}a, the sharp changes in the cost function occur during the iterations, as the Q-FEM problem via the VQLS algorithm is a non-convex optimization problem, as discussed previously. Figure \ref{fig:cost_function_results_quadratic_elements}b shows that the Q-FEM solution matches with the classical FEM solution for both $n=3$ and $n=4$ examples. The 3-qubit and 4-qubit results are different in amplitude because they are normalized with respect to the size of the solution vector.

\begin{table}[htb!]
    \small
    \centering
    \begin{tabular}{|c|c|c|c|}
    \hline 
     \# of qubits & Circuit depth & One-qubit gates & Two-qubit gates\\ \hline
    3 & 352 & 262 & 212 \\  
    4 & 1036 & 737 & 518 \\ 
    \hline
    3 & 325 & 257 & 188 \\
    4 & 883 & 630 & 488 \\  
    \hline
    \end{tabular}
    \caption{The maximum numbers of total circuit depth, one-qubit, and two-qubit gate count among all circuits (transpiled) for problem in Fig.~\mbox{\ref{fig:cost_function_results_quadratic_elements}}. The second and third rows correspond to $ \langle \bs{\psi}| \bs{\psi} \rangle$ circuits, while the last two rows correspond to $ \langle \bs{\psi}| f \rangle$ circuits.}
    \label{tab:circuit_characteristics_quadratic}
\end{table}

Table~\mbox{\ref{tab:circuit_characteristics_quadratic}} reports the maximum number of circuit depth, one-qubit, and two-qubit gate counts for the quadratic elements problem given in Fig.~\mbox{\ref{fig:cost_function_results_quadratic_elements}}, among all possible $ \langle \bs{\psi}| \bs{\psi} \rangle$ and $\langle \bs{\psi}| f \rangle$ circuits. Both the circuits are transpiled including the Hadamard gates (for measurement) and variational ansatz gates with appropriate circuit depth as stated above, and the basis gates and optimization level chosen for tranpilation is the same as reported earlier in Sec.~\mbox{\ref{sec:complexity}}. 

\vspace{-0.3cm}
\section{Conclusions}
\label{section:Conclusion}
In this work, the Q-FEM framework was developed for use in NISQ computers and implemented using the VQLS algorithm on IBM's \emph{Qiskit} simulator. The framework keeps the FEM discretization of the domain intact, which allows the use of variable element lengths and material coefficients, for both linear and quadratic element discretization. The framework is applied to the steady-state heat equation, and a general formalism is developed to generate the quantum circuits corresponding to the unitary decomposition of the stiffness matrix. It is shown through various examples for both linear and quadratic elements discretization that the Q-FEM framework reproduces the classical FEM results, with variable element lengths and non-homogeneous material coefficients. The Q-FEM framework is an explicit scheme representing the FEM stiffness matrix with quantum circuits, and is not limited to use in variational solvers, and could also be applied with other quantum linear solvers.

Several advancements are still needed to leverage the benefits of quantum computing in solving computational mechanics problems. A key difficulty is achieving scalability given the convergence difficulties in the VQLS algorithm. The degradation of the stiffness matrix conditioning with increasing mesh density results in convergence difficulties primarily due to barren plateaus and the solutions being `hidden' in narrow and deep gorges. New algorithms that can perform efficiently in the presence of badly conditioned problems are therefore necessary; although, improved performance could be expected from solving preconditioned linear system, informed ansatz design, and using optimization methods such as differential evolution algorithms. Furthermore, the current manuscript focused on implementing the generator function approach to compute the unitary matrices for 1-D problems. A non-trivial extension of the generator function approach will be required for multidimensional discretizations to solve more complex problems using the proposed approach. Lastly, future investigations will focus on the verification of the proposed approach in real quantum devices and the assessment of Q-FEM's robustness against errors that exist in the current NISQ devices.

\vspace{-0.3cm}
\section*{Acknowledgments}

The authors gratefully acknowledge the funding support from the National Science Foundation, CMMI Division, Mechanics of Materials and Structures Program (Award No:2222404) and Vanderbilt University Seeding Success Program.

\appendix

{\bibliography{References}}

\end{document}